\documentclass{aa}  

\usepackage{graphicx}

\usepackage{booktabs}
\usepackage{xcolor}
\usepackage{ulem} 
\usepackage{natbib}
\bibpunct{(}{)}{;}{a}{}{,} 
\usepackage{lineno}
\usepackage{txfonts}

\newcommand{\hi}{{\rm H}{\textsc i}}
\newcommand{\hii}{{\rm H}{\textsc{ii}}}

\usepackage[T1]{fontenc}
\usepackage[colorlinks=true,linkcolor=black, urlcolor=black, citecolor=black]{hyperref}

\begin{document}

   \title{Unveiling the nature and fate of the almost-dark cloud AGC 226178 through $\hi$ mapping}

   \subtitle{}

   \author{Yu-Zhu Sun
          \inst{1,2,3}
          \and
          Hong-Xin Zhang\inst{1,2}\fnmsep\thanks{Corresponding author: hzhang18@ustc.edu.cn}
          \and 
          Elias Brinks\inst{4}
          \and 
          Rory Smith\inst{5,6}
          \and 
          Fujia Li\inst{1,2}
          \and
          Minsu Kim\inst{7,8}
          \and
          Se-Heon Oh\inst{7,8}
          \and
          Zesen Lin\inst{9}
          \and
          Jaebeom Kim\inst{7,8}
          \and
          Weibin Sun\inst{1,2}
          \and
          Tie Li\inst{1,2}
          \and
          Patrick C\^ot\'e\inst{10}
          \and
          Alessandro Boselli\inst{11}
          \and 
          Lijun Chen\inst{1,2}          
          \and
          Pierre-Alain Duc\inst{12}
          \and
          Sanjaya Paudel\inst{13}
          \and
          Matthew A. Taylor\inst{14}
          \and 
          Kaixiang Wang\inst{15,16} 
          \and
          Enci Wang\inst{1,2}
          \and
          Lanyue Zhang\inst{1,2}
          \and
          Yinghe Zhao\inst{17,18}
          }

   \institute{Department of Astronomy, University of Science and Technology of China, Hefei 230026, China
         \and
         School of Astronomy and Space Science, University of Science and Technology of China, Hefei 230026, China
         \and
             School of Physics and Astronomy, University of Leicester, University Road, Leicester, LE1 7RH, UK
         \and
             Centre for Astrophysics Research, University of Hertfordshire, College Lane, Hatfield, AL10 9AB, UK
         \and
         Departamento de Fisica, Universidad Tecnica Federico Santa Maria, Avenida España 1680, Valparaíso, Chile
         \and
         Millenium Nucleus for Galaxies (MINGAL)
         \and
         Department of Astronomy and Space Science, Graduate School, Sejong University, 209 Neungdong-ro, Gwangjin-gu, Seoul, Republic of Korea
         \and
         Department of Physics and Astronomy, Sejong University, 209 Neungdong-ro, Gwangjin-gu, Seoul, Republic of Korea
         \and
         Department of Physics, The Chinese University of Hong Kong, Shatin, New Territories, Hong Kong SAR, China
         \and
         National Research Council of Canada, Herzberg Astronomy and Astrophysics Research Centre, Victoria, BC V9E 2E7, Canada
         \and
         Aix Marseille Univ, CNRS, CNES, LAM, Marseille, France
         \and
         Université de Strasbourg, CNRS, Observatoire astronomique de Strasbourg, UMR 7550, Strasbourg, France.
         \and 
         Department of Astronomy \& Center for Galaxy Evolution Research, Yonsei University, Seoul 03722, Republic Of Korea
         \and
         University of Calgary, 2500 University Drive NW, Calgary, Alberta, T2N 1N4, Canada
         \and
         Department of Astronomy, Peking University, Beijing, China
         \and
         Kavli Institute for Astronomy and Astrophysics, Peking University, Beijing, China 
         \and
         Yunnan Observatories, Chinese Academy of Sciences, Kunming 652016, People's Republic of China
         \and
         State Key Laboratory of Radio Astronomy and Technology, National Astronomical Observatories, Chinese Academy of Sciences, Beijing 100101, China
             }

   \date{Received ; accepted }

  \abstract
  {The origin of extragalactic, almost dark $\hi$ clouds with extreme gas-to-stellar mass ratios remains poorly understood.}  
  {We  investigated the nature and fate of the almost dark cloud AGC 226178, projected within the Virgo cluster, which exhibits an $\hi$-to-stellar mass ratio of approximately 1000.}
  {We present deep single-dish $\hi$ mapping from the Five-hundred-meter Aperture Spherical Telescope (FAST), complemented by  high-resolution interferometric data from the Very Large Array (VLA), as part of the Atomic gas in Virgo Interacting Dwarf galaxies (AVID) project. Together, these observations provide the highest-quality $\hi$ analysis to date of an almost dark cloud, in terms of the combination of spatial resolution and sensitivity.}
  {The FAST data reveal a short low-velocity tail extending  towards the dwarf galaxy VCC~2034, which was previously proposed as a possible origin for AGC~226178. However, VCC~2034 itself exhibits a line-of-sight asymmetric $\hi$ feature and a cometary morphology that indicates a stripping event unrelated to AGC 226178. 
  VLA observations reveal a velocity gradient across AGC~226178, along with a clumpy internal structure. The velocity dispersion within the cloud exceeds the thermal line width, indicating the presence of turbulence and/or unresolved random motions. The whole cloud cannot be gravitationally bound by the observed atomic gas alone. The resolved $\hi$ clumps follow the standard $\hi$ mass–star formation rate relation and a tight mass–size relation:  those associated with star formation reach surface densities above the theoretical threshold for self-shielding.}
  {We conclude that AGC~226178 is a free-floating $\hi$ cloud of unknown origin. The system appears to be in the process of disintegration. It is likely located well outside the Virgo cluster, as the preservation of its extended $\hi$ morphology within the cluster environment would otherwise require a substantial reservoir of unseen molecular gas with a mass exceeding that of the observed $\hi$ content. While confinement pressure from the hot intracluster medium may contribute to its stability, it is unlikely to be the dominant factor preventing its disruption.
  }


   \maketitle

\section{Introduction} \label{sec:intro}

Over the years, several objects have been detected which have been proposed as optically dark galaxy candidates. With recent blind $\hi$ surveys, such as the Arecibo Legacy Fast ALFA blind $\hi$ survey \citep[ALFALFA,][]{2005AJ....130.2598G}, the FAST all sky HI survey \citep[FASHI,][]{2024SCPMA..6719511Zhang}, and the Widefield ASKAP L-band Legacy All-sky Blind surveY\citep[WALLABY,][]{2020Ap&SS.365..118K}, gas-rich objects with extremely faint or even no optical counterpart have been revealed. According to cosmological simulations in the prevalent $\rm \Lambda CDM$ paradigm, the expected number of dark matter subhalos  around a Milky Way-size host galaxy is more than an order of magnitude larger than the detected satellite galaxies, which is known as the `missing satellite' problem \citep[][]{Klypin1999}. Either the standard CDM model is fundamentally flawed, or more low-mass halos would remain starless or extremely inefficient in star formation and thus still awaiting detection. In addition, different dark matter models predict different subhalo number distribution in the low-mass end \citep[e.g. warm dark matter models;][]{2001ApJ...556...93B}. Furthermore, small dark halos with inefficient or no star formation, and thus weak baryonic feedback effects, are expected to retain the original dark matter halo structure. Therefore, identifying (almost) optically dark dwarf galaxies from observations may lead to crucial tests for both dark matter and galaxy formation models. 

Dark matter halos at the  lower-mass end are more prone to baryon depletion, which means they could be hard to observe. Below a characteristic halo mass, galaxy star formation may be completely suppressed due to severe baryon depletion and inefficient gas cooling \citep{Hoeft2006}, which is partly attributed to photoheating by the cosmological ultraviolet (UV) background. \cite{2010AdAst2010E..87H} find the characteristic halo mass to be $\sim 6\times 10^{9}h^{-1}M_{\odot}$, which is robust against the assumed UV background flux density and resolution effects in their simulations. In addition, stellar feedback could also remove the gas component from galaxies with shallow potential wells.
While it is generally challenging to detect stellar light from almost dark galaxies, it is   easier to detect them through neutral $\hi$ gas emission \citep[e.g.][]{2006MNRAS.368.1479Davies}. 

Several candidates for (almost) dark galaxies have been reported in the past few decades:  VIRGOHI21 \citep[][]{Davies2004}, Nube \citep[][]{Montes2024}, FAST J1039+4328 \citep{2023ApJ...944L..40Xu}, and H1032-2819 \citep{2024MNRAS.528.4010OBeirne}. 
To mention a few of the candidate dark galaxies, FAST J1039+4328 \citep{2023ApJ...944L..40Xu} shows a double-horned $\hi$ velocity profile, which is in line with a disc in regular rotation.
\cite{2015ApJ...801...96Janowiecki} reported high-resolution observations of the isolated (no known neighbouring galaxies closer than $\rm \sim 500\ kpc$) $\hi$ system HI1232+20 (which consists of three almost dark clouds: AGC 229383, AGC 229384, and AGC 229385). The main $\hi$ part of the system has a faint optical counterpart with a $g$-band peak surface brightness of $\rm \sim 26.4\, mag\ arcsec^{-2}$. The two fainter gas clouds have no optical counterpart down to 27.9 and 27.8 mag $\rm arcsec^{-2}$ in the $g$-band \citep{2015ApJ...801...96Janowiecki}.
The nature of the system is still controversial; it lacks tidal features and a clear connection to other galaxies in the local environment. 

VIRGOHI21 was originally postulated to be a true dark galaxy \citep{2005ApJ...622L..21Minchin,2007ApJ...670.1056Minchin}; however, subsequent observations \citep{2007ApJ...665L..19Haynes} revealed its elongated nature and faint gas connection with a larger galaxy. Simulations \citep{2008ApJ...673..787Duc} have also shown that this source is likely a tidal remnant. The dark cloud AGC 229101 has no optical counterpart in the SDSS imaging survey, but deeper observations \citep{2021AJ....162..274Leisman} revealed a faint stellar component. Higher-resolution $\hi$ mapping indicates that it may be explained by a tidal encounter with neighbouring objects and a merger of two dark $\hi$ clouds. 

The origin of most of the observed dark galaxy candidates is still shrouded in mystery. Although highly elongated $\hi$ structures are better explained as remnants of tidal interactions \citep[e.g.][]{2008ApJ...673..787Duc,2021ApJ...922L..21Zhu,2024MNRAS.532.1744Yu} or ram pressure stripping \citep[e.g.][]{2024A&A...690A...4Serra}, the origin of apparently roundish and isolated $\hi$ clouds is challenging our understanding: they genuinely could be dark galaxies. Beyond their potential importance in understanding the nature of dark matter, (almost) dark galaxy candidates are also valuable for studying the star formation process in an extremely $\hi$-rich environment.

AGC 226178 is a gas-rich, faint blue stellar system in the Virgo cluster, first identified by the ALFALFA survey and classified as an `Almost Dark' object \citep{2008AJ....136..713Kent,2015AJ....149...72Cannon}. The system contains approximately $\rm 4 \times 10^7\ M_{\odot}$ of $\hi$, with a line-of-sight velocity of $\rm 1581 \pm 6\ km\ s^{-1}$, consistent with being located near the Virgo cluster.

Follow-up observations conducted with the NRAO\footnote{The National Radio Astronomy Observatory is a facility of the National Science Foundation operated under cooperative agreement by Associated Universities, Inc.} Karl G. Jansky Very Large Array (VLA) in D-configuration by \cite{2015AJ....149...72Cannon} revealed that the neutral hydrogen in this system is centrally concentrated, exhibiting a well-defined velocity gradient spanning approximately $\rm 10\ km\ s^{-1}$. Based on the excellent positional agreement between the $\hi$ centre and the UV-bright stellar component, along with the inclined disc morphology observed in the UV emission, they concluded that AGC 226178 is likely a low-mass dwarf galaxy.

With deep optical observations from the Next Generation Virgo cluster Survey \citep[NGVS,][]{2012ApJS..200....4Ferrarese}, \cite{2021A&A...650A..99Junais} estimated the stellar mass of AGC 226178 as $\rm \sim 5\times 10^4M_{\odot}$.\ They 
brought up the idea that AGC 226178  resulted from ram pressure acting on the nearby ultra-diffuse galaxy (UDG) NGVS 3543. This scenario might explain the faint bridge or clump of UV emission projected between the two objects (and also to the south of AGC 226178). Nevertheless, AGC 226178 and NGVS 3543 were later resolved into individual stars in Hubble Space Telescope (HST) imaging presented in \cite{2022ApJ...926L..15Jones}. These authors found that NGVS 3543 is likely a foreground dwarf galaxy at a distance of $\sim$ 10 Mpc. Moreover, the metallicity expected for the mass of NGVS 3543 is more than an order of magnitude lower than the measured nebular gas metallicity of AGC 226178. A re-analysis of the ALFALFA data by \cite{2022ApJ...926L..15Jones} revealed an extremely faint, bridge-like structure between AGC 226178 and the dwarf irregular galaxy VCC 2034. Considering the cluster environment in which the two objects reside, they concluded that AGC 226178 likely originated from VCC 2034 through a ram pressure stripping event. 

The origin of AGC 226178 remains inconclusive. Deeper and higher-resolution mapping of the $\hi$ gas is necessary to help resolve the ambiguity. AGC 226178 was recently observed by the Atomic gas in the Virgo Interacting Dwarf galaxies (AVID, Zhang et al. in prep.) project, which obtained both multi-configuration VLA observations and deep FAST mapping of the $\hi$ emission line for a sample of dwarf galaxies involved in tidal interactions or dwarf-dwarf galaxy mergers. AGC 226178 was not among the core sample of AVID, but it falls into one of the AVID fields. For this paper we performed a joint analysis of the AVID observations and ancillary multi-frequency data of AGC 226178 in order to shed further light on the origin and evolution of this intriguing, almost dark object.

\section{Observations and data reduction} \label{sec:style}
\subsection{VLA $\hi$ data}

In this work, we use $\hi$ 21-cm emission line observations of AGC 226178 obtained with the VLA. The data consist of B, C and D multi-configuration observations from the AVID project (PID: 21A-231; PI: Hong-Xin Zhang), and the archival D-configuration data first presented by \cite{2015AJ....149...72Cannon}.

We observed the $\hi$ 21-cm line of AGC 226178 in the L-band centred at 1413.96 MHz, with a subband bandwidth of 16 MHz and channel width of $\rm 7.81\ kHz\ (\sim 1.67\ km\ s^{-1})$. The AVID D configuration observation was carried out on April 9, 2021 for 1.5 hr (1 hr on source), the C configuration observation on June 21, 2021 for 2.5 hr (2 hr on source), and the B configuration observations on November 10, December 16, and December 17 of 2021 for a total of 6 hr (4.7 hr on source). 3C 286 was observed as the flux and bandpass calibrator at the beginning and end of every execution block. The complex gain calibrator, J1254+1141, was observed for effectively 3 min, for every 15-17 minutes on the target field. 

The VLA data calibration and imaging were performed with the Common Astronomy Software Applications  (\verb|CASA|) \citep{2007ASPC..376..127McMullin} version 6.5.4-9. Particularly, the flux and phase calibration was performed primarily through the VLA Calibration Pipeline (adapted for emission line), with additional extensive manual flagging of $uv-$data affected by radio frequency interference (RFI) by using the \verb|PLOTMS| and \verb|FLAGDATA| tasks in \verb|CASA|. Hanning smoothing was applied to the $uv-$data to yield an effective spectral resolution of $\rm 3.4\ km\ s^{-1}$. We performed continuum subtraction using the task \verb|UVCONTSUB|. Each execution block was calibrated and continuum-subtracted separately.

AGC 226178 was previously observed as part of the ALFALFA ‘almost dark’ galaxies sample \citep{2015AJ....149...72Cannon} in VLA D configuration. We reduced this dataset following the same procedure as our own observations. The data was collected with the same channel width as our observations and a total bandwidth of 8 MHz.

All of the above calibrated and continuum-subtracted multi-configuration datasets were combined and imaged with \verb|tclean|. Two $uv$ weighting methods were used for imaging. One method is natural weighting; the other is  Briggs' weighting with a robust parameter set at $r$=0.5 (hereafter robust weighting). The natural weighting produces an image with the lowest noise, but at the expense of somewhat lower angular (and thus spatial) resolution, whereas the robust weighting attempts to balance the sensitivity and resolution. In addition to the above standard weighting schemes, we also applied a $uv$ taper of 
20$k\lambda$
to the natural weighting scheme to reach an even higher column density sensitivity on large scales, at the expense of an even lower spatial resolution.
To produce the imaging cubes, we run the \verb|tclean| task for each weighting scheme, applying a circular clean box centred on the source with a radius of $4\arcmin$. The deconvolution was carried out down to a $2$-$\sigma$ noise level.

The resulting beam size is 
$\rm 7.15^{\prime\prime}\times6.45^{\prime\prime}$ for the robust-weighted image cube, $\rm 14.48^{\prime\prime}\times13.38^{\prime\prime}$ for the natural-weighted image cube and $\rm 20.5^{\prime\prime}\times 19.5^{\prime \prime}$ for the $uv$-tapered natural weighted cube. The resulting root mean square  (RMS) noise is $\rm 0.47\ mJy\,beam^{-1}$, $\rm 0.40\ mJy\,beam^{-1}$, and $\rm 0.47\ mJy\,beam^{-1}$ for the robust-weighted, natural-weighted, and $uv$-tapered image cubes, respectively. This corresponds to a minimum detectable $\hi$ column density of $\rm 2.0\times 10^{20}\ cm^{-2}$, $\rm 3.9\times 10^{19}\ cm^{-2}$, and $\rm 2.2\times 10^{19}\ cm^{-2}$ (3$\sigma$ over $\rm 20\ km\ s^{-1}$), respectively.

The final source masks used for generating moment maps are also generated by \verb|SoFiA 2|, with the S+C spatial smoothing kernels set to 0 and 1.5 times the synthesized beam size, the spectral smoothing kernels to 0 and 3 channels, and a detection threshold of 3-$\sigma$ over each smoothing scale. We limited the first moment (intensity-weighted mean velocity field) and second moment (velocity dispersion map) maps to only include the pixels where the signal-to-noise ratio is larger than 3. 
A more thorough description of the data reduction and emission detection will be presented in Zhang et al.(in prep.)

\subsection{FAST $\hi$ data}

We obtained deep $\hi$ observations of AGC 226178 with FAST as part of the AVID project (PID: PT2022-0005; PI: Hong-Xin Zhang).
FAST is the largest single-dish radio telescope in the world, with a 500 m diameter aperture (300 m illuminated aperture) \citep{Nan2011,Jiang2019,Qian2020}.
Its L-band 19-beam receivers have a frequency range of 1.05-1.45 GHz in four polarization modes and the half-power beam width (HPBW) of each beam is $\sim$ 2.90$^\prime$ at 1.4 GHz (3.24$^\prime$ after gridding in imaging process).
We used the Multibeam on-the-fly (OTF) mode to uniformly map a $\sim$ 30\arcmin $\times$ 30\arcmin~sky area centred on VCC2037. AGC 226178 is about 12\arcmin~to the north-east of the field centre. Every observing block consists of a pair of horizontal and vertical scans, with a 23.4$^\circ$/53.4$^\circ$ (horizontal/vertical) rotation of the receiver, a 21.66$^\prime$ scan separation, and 15$\rm ^{\prime \prime} s^{-1}$ scan speed. 
We followed the observing strategy described in \citet{2023ApJ...944..102Wang} and \citet{2025ApJ...984...15Yang}.
The Spec(W+N) backend was used to record the data with a sampling time of 0.5 s, covering 500 MHz bandwidth with 65536 channels. The frequency and velocity resolution are 7.63 kHz and 1.61 km s$^{-1}$ at 1.42 GHz.
The flux calibration was accompanied by injecting a 10 K noise diode for 1s every 60s on target in order to calibrate the antenna temperature.

The observations were carried out on August 13, 21, 23, and September 4 of 2022. The accumulated on-source time per sky position is about 4.87 minutes.
The raw data were reduced mainly using the $\hi$FAST pipeline\footnote{\url{https://hifast.readthedocs.io/en/v1.3/}} developed by \citet{2024SCPMA..6759514Jing}. The pipeline follows the general reduction procedure for single-dish data. With $\hi$FAST, we first chose the FFT-filter approach to remove standing waves (more details can be found in \citealt{2025RAA....25a5011X}).
Then we used the \texttt{MedMed} algorithm with the \texttt{poly-asym2} method to perform the post-processing for baseline subtraction. After performing RFI flagging, frame conversion, and velocity correction, the final calibrated spectra were gridded into a data cube with a pixel size of 1$^\prime$. The data cube was further Hanning smoothed to a velocity resolution of 6.4 km s$^{-1}$.
The final FAST data cube has an RMS noise level of 0.78 $\rm mJy\,beam^{-1}$ prior to Hanning smoothing, which corresponds to a minimum detectable $\hi$ column density of $\rm 4.8\times 10^{17} cm^{-2}$  (3$\sigma$ over 20 $\rm km\ s^{-1}$).

As with the VLA data, we used the S+C algorithm in \verb|SoFiA 2| to search for genuine $\hi$ emission signal in the FAST image cube. Particularly, we chose spatial smoothing kernels of 0, 1 and 2 times the beam size, spectral smoothing kernels of 0, 5 and 12 $\rm km\,s^{-1}$, and a detection threshold of $3.5\sigma$ for each smoothing scale. We searched for emission over a Heliocentric velocity range of $\rm 500\sim 3000 km\ s^{-1}$.

\section{Analysis and results} 

\subsection{Large-scale $\hi$ distribution}\label{sec: lscalehi}

\begin{figure*}[htp]
   \centering
   \includegraphics[width=18cm]{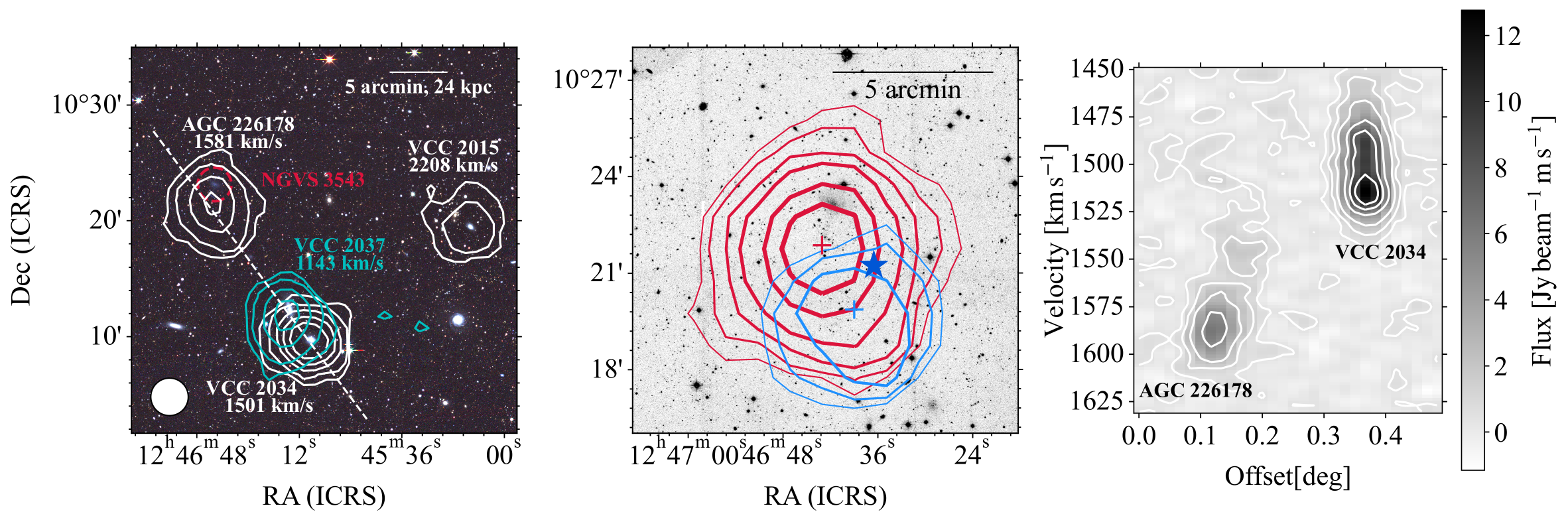}
   \caption{\textit{Left}: DECaLS \textit{g,r,z} composite colour image overlaid with integrated $\hi$ emission (moment 0 map) contours from the FAST observation. The contour levels are (0.3, 1.1, 4.2, 8.4, 16.8)$\rm \times10^{18}\,cm^{-2}$. The beam is shown as the white circle in the bottom left corner ($\rm 3.24\arcmin \times 3.24\arcmin$). The scale bar in the top right corner illustrates the angular size and its corresponding physical scale, assuming a distance of 16.5 Mpc. The white dashed line running through AGC 226178 represent the directions adopted to extract the position-velocity diagrams shown in the right panels. \textit{Middle:} Zoomed-image of  NGVS $g$-band  centred on AGC 226178 overlaid with FAST contours, red for the main body, blue for the tail of AGC 226178. The contour levels are (0.3, 0.5, 1.1, 2.1, 4.2, 6.3)$\rm \times10^{18}\,cm^{-2}$. The position of BC12 is indicated by the blue star. \textit{Right}: FAST $\hi$ position–velocity diagram extracted along the dashed line shown in the left panel. The extraction slit is 3\arcmin\ wide (centred on the dashed line), matching the beam size of the FAST resolution. The x-axis indicates the offset along the dashed line, measured from the reference position (RA, Dec) = $12^{\mathrm{h}}47^{\mathrm{m}}03\fs38$, $10^\circ27\arcmin44\farcs83$. The contours are drawn at 0.4, 1.2, 2.5, 5.0, 7.4, and 9.9 $\rm mJy\,beam^{-1}$, approximately $1\sigma$, $3\sigma$, $6\sigma$, $12\sigma$, $18\sigma$, and $24\sigma$ (over a single channel which corresponds to 1.61 $\rm km\ s^{-1}$). VCC 2037 ($\rm \sim 1150~km\ s^{-1}$) does not appear in the diagram as it is well beyond the plotted velocity range.}
   \label{fig: DESI_FAST_PV}
\end{figure*}

\begin{figure}
\includegraphics[width=9cm]{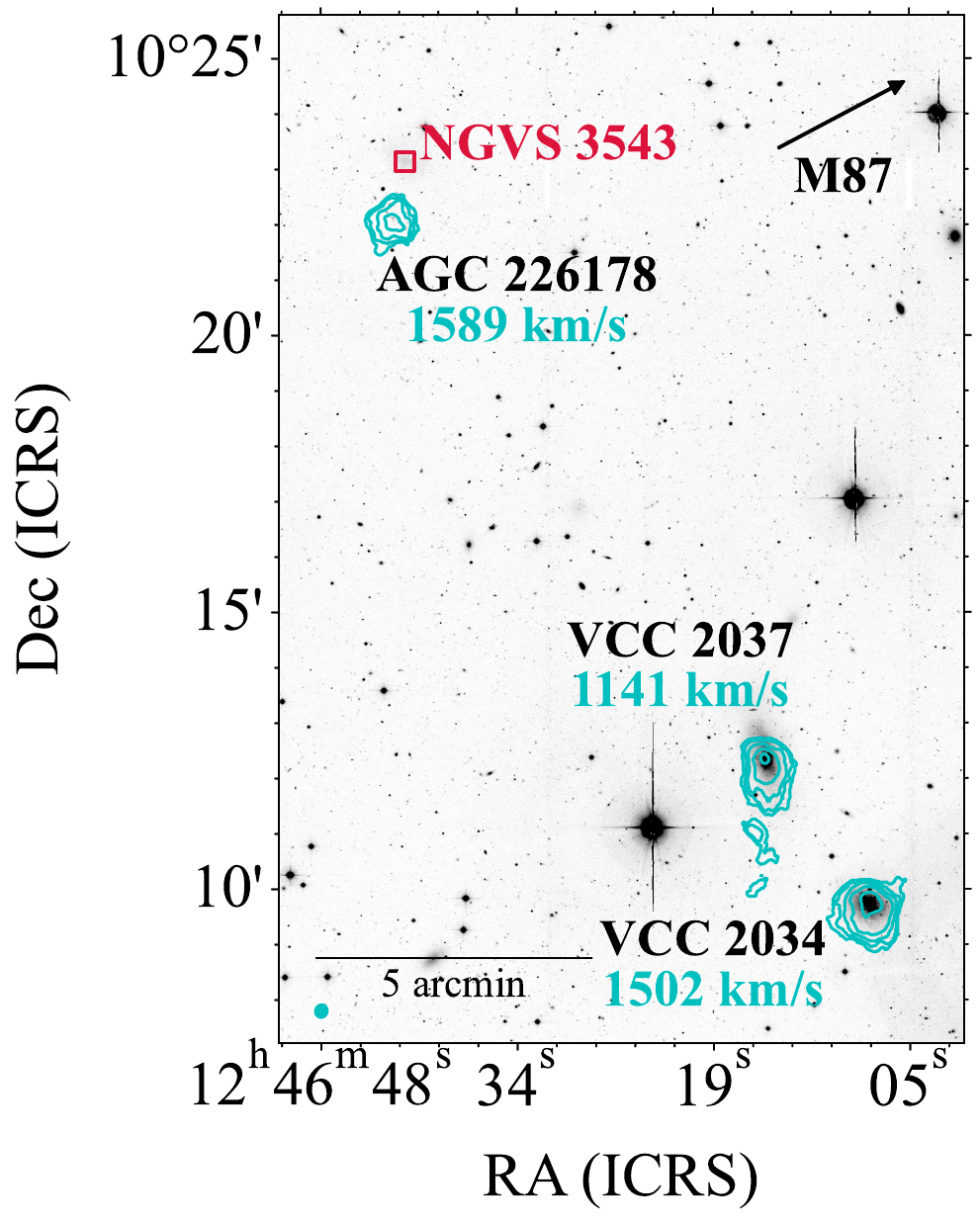}
\caption{Natural-weighted VLA moment-0 map showing all detected sources in the field, including AGC 226178, VCC2034, and VCC~2037. Cyan contours represent column density of (0.4, 0.8, 1.6, 3.2, 6.4)$\rm\times10^{20}\,cm^{-2}$. The central velocity of each source was measured from the masked VLA data cube. The location of NGVS 3543, which is not detected in the VLA data, is indicated by the crimson rectangle. The arrow indicates the direction towards M87, and the cyan ellipse in the lower left corner represents the beam size of the moment maps. }
\label{fig: vla_field}
\end{figure}

Fig.\ref{fig: DESI_FAST_PV} shows the FAST $\hi$ column density contours of the AGC 226178 field, overlaid on the Dark Energy Camera Legacy Survey (DECaLS) false-colour image \citep{2019AJ....157..168DESI}. The 3$\sigma$ sensitivity for a 20 $\rm km\ s^{-1}$ line width of our FAST map is $\rm\ 4.8\times 10^{17} cm^{-2}$, which is approximately one-sixth of the detection limit of ALFALFA observations \citep[$\rm 2.8\times10^{18}\,cm^{-2}$,][]{2022ApJ...926L..15Jones}. Such a substantial improvement in sensitivity is expected to significantly enhance the ability to detect faint $\hi$ features on large scales. Complementarily, Fig. \ref{fig: vla_field} displays the natural-weighted moments 0 contours from VLA observations, overlaid on the NGVS $g$-band image, providing the highest spatial resolution available for this field.

All of the galaxies detected in the FAST field, as indicated in Fig.\ref{fig: DESI_FAST_PV}, are dwarf galaxies with relatively small angular sizes and barely resolved by FAST.  
Some pertinent $\hi$ properties of these objects (AGC 226178, VCC 2034, VCC 2037 and VCC 2015) are listed in Table \ref{tab:fast}. We note that the blue compact dwarf galaxy VCC 2015 \citep[$(\alpha, \delta) = (12^h 45^m 13^s.30, 10^d19^m33^s.23)$;][]{1987ApJS...63..247Hoffman,2005A&A...429..439Gavazzi} is detected in $\hi$ emission for the first time here, with an integrated $\hi$ flux of $\rm S=0.35\pm0.04\ Jy\ km\ s^{-1}$ and a systemic Heliocentric velocity of $\rm V_{sys}=2208.1\ km\ s^{-1}$. Assuming the average distance of the Virgo cluster (16.5 Mpc) \citep{2024ApJ...966..145Cantiello}, the corresponding $\hi$ mass is $\rm log(M_{HI}/M_{\odot})=7.35\pm 0.10$. 

The dwarf galaxies VCC 2037 and VCC 2034 cannot be spatially distinguished in the FAST integrated column density map, due to their small angular separation. But they are well-separated in velocity, with a difference of $\rm \sim 400\ km\ s^{-1}$ in systemic velocity. AGC 226178 appears more extended towards the direction of VCC 2034 than in the opposite direction. As indicated in the position-velocity (PV) diagrams along the AGC 226178-VCC 2034 direction (dashed line in the left panel) shown in the right panel of Fig.\ref{fig: DESI_FAST_PV}, AGC 226178 exhibits a velocity gradient along the direction, particularly with a faint, extended component towards the lower-velocity end. 

The overall $\hi$ morphology of AGC 226178 from FAST shows a compact main body and an extended, kinematically distinct tail component with central radial velocity of $\sim$ 1550 km s$^{-1}$, consistent with a head-tail structure expected from ongoing stripping. The two components in the moment-0 map are integrated over the velocity ranges 1535–1565 $\rm km\,s^{-1}$ for the tail, and 1565–1625 $\rm km\,s^{-1}$ for the main body (see middle panel of Fig.\ref{fig: DESI_FAST_PV}). 
The angular separation between the flux centres of main body and tail is measured to be 2.24 arcminutes. Assuming the Virgo cluster distance, this corresponds to a projected physical separation of 10.7 kpc. 

In a recent study, \citet{2025ApJ...983....2Dey} discovered another Blue Cloud (BC12) near AGC 226178 (referred to as BC3 in their paper), at a projected distance of only 9 kpc. The location of BC12 is marked with a blue star in the middle panel of Fig.~\ref{fig: DESI_FAST_PV}. Observations of the H$\alpha$ emission line indicate that the velocity of BC12 is consistent with that of AGC 226178, within the limits of the spectral resolution and observational uncertainties. The two systems also exhibit comparable metallicities. Moreover, the stellar component of BC12 appears to lie along the extension of the elongated optical structure of BC3. Therefore, it is highly likely that there was a physical connection between these two objects. Nevertheless, the position of BC12 is clearly offset from the direction of the $\hi$ tail detected in our FAST observations.

BC12 is not detected in our VLA $\hi$ images, we therefore place a 3$\sigma$ upper limit on the neutral $\hi$ content of BC12 of $\rm \sim 2\times10^6\,M_\odot$, by adopting a 15\arcsec-diameter circular aperture and a 10.2 km/s linewidth for the tapered natural-weighted image cube.

\begin{figure}
\includegraphics[width=9cm]{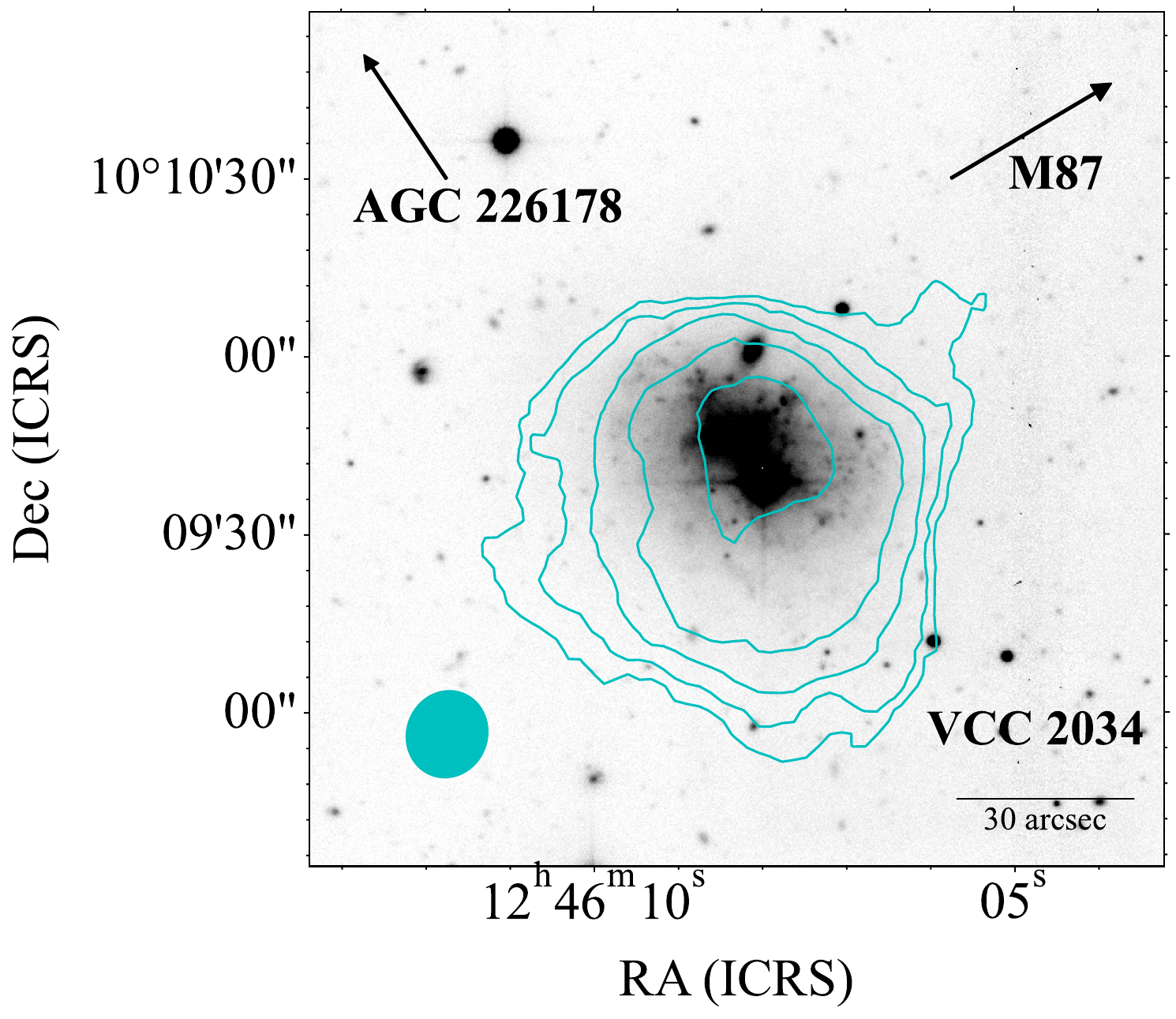}
\caption{Zoomed-in view of the natural-weighted VLA moment-0 map of VCC 2034. The two arrows indicate the directions towards AGC 226178 and the centre of the Virgo cluster, respectively. The contours follow the same levels as those in Fig.~\ref{fig: vla_field}.}
\label{fig: vla_VCC2034}
\end{figure}

We note that \cite{2022ApJ...926L..15Jones} found a low signal-to-noise $\hi$ connection between AGC 226178 and VCC 2034, based on the ALFALFA observation. Our deeper FAST observation reveals a short extension towards VCC 2034, but disproves a direct connection between the two. Moreover, the right PV diagram in Fig.\ref{fig: DESI_FAST_PV} also reveals an asymmetric extension in the line-of-sight velocity distribution towards the lower-velocity end of VCC 2034. In contrast to the lower-velocity end, the contours at the higher-velocity end of VCC 2034 appear relatively squeezed, which is an indication of ongoing stripping. The stripping direction in velocity space points away from, rather than towards, AGC 226178. Although VCC 2034 is not spatially resolved in the FAST observation, the VLA image reveals a cometary morphology, with the head orientated roughly towards the north, instead of along the direction to AGC 226178 (Figs. \ref{fig: vla_field} and \ref{fig: vla_VCC2034}). Therefore, both the $\hi$ morphology and the radial velocity—along with the direction of its gradient—of the $\hi$ stripping features argue against the possibility that AGC 226178 originated from VCC 2034. The misalignment of the stripping tails in both systems with respect to the direction towards the Virgo cluster centre suggests that they possess significant tangential motion within the cluster, rather than undergoing purely radial infall. 

The angular separation between AGC 226178 and its low surface brightness neighbour NGVS 3543 is only 1 arcmin; hence, they cannot be spatially distinguished in the FAST data. NGVS 3543 does not yet have an $\hi$ detection, but according to the model predictions by \cite{2022A&A...667A..76Junais}, if AGC 226178 originated from a ram pressure stripping event of NGVS 3543, then NGVS 3543 should still possess gas with a mass of $\rm \sim 10^5 \ M_{\odot}$. Therefore, we attempted to search for such a faint $\hi$ source using our data. \cite{2022A&A...667A..76Junais} estimated the stellar mass of NGVS 3543 $\log M_*/M_{\odot}=7.10\pm 0.02$. According to the Baryonic Tully-Fisher relation (BTFR) from \cite{2016A&A...593A..39Papastergis}, the possible maximum rotation velocity of NGVS 3543 is $v_{rot}$  $\rm \simeq 20\ km\ s^{-1}$. Assuming the $\hi$ full width at 50\% of the peak intensity $W_{50}=2v_{rot}$ \citep[][]{2021MNRAS.508.1195Ponomareva}, we expect to detect an $\hi$ mass down to $\sim 6.9\times10^5\ M_{\odot}$ with a significance of $5\sigma$. If NGVS 3543 contains $\hi$ gas as expected by \cite{2022A&A...667A..76Junais}, then it would be near the detection limit of the FAST observation. According to the Systematically Measuring Ultra-diffuse Galaxies (SMUDGes) survey \citep{2023ApJS..267...27Zaritsky_ngvs}, the approximate velocity of NGVS3543 is reported as $\rm 1511\,km\,s^{-1}$, estimated using the Distance-by-Association method. This approach assigns a redshift to ultra-diffuse galaxies based on the median redshift of at least three bright galaxies located within a projected distance of 1.5 Mpc. However, such estimates are highly uncertain for nearby systems, particularly within the Virgo cluster. While we took this velocity into consideration, we also searched for emission over a broader velocity range.

\begin{figure*}[htp]
   \centering
   \includegraphics[width=16cm]{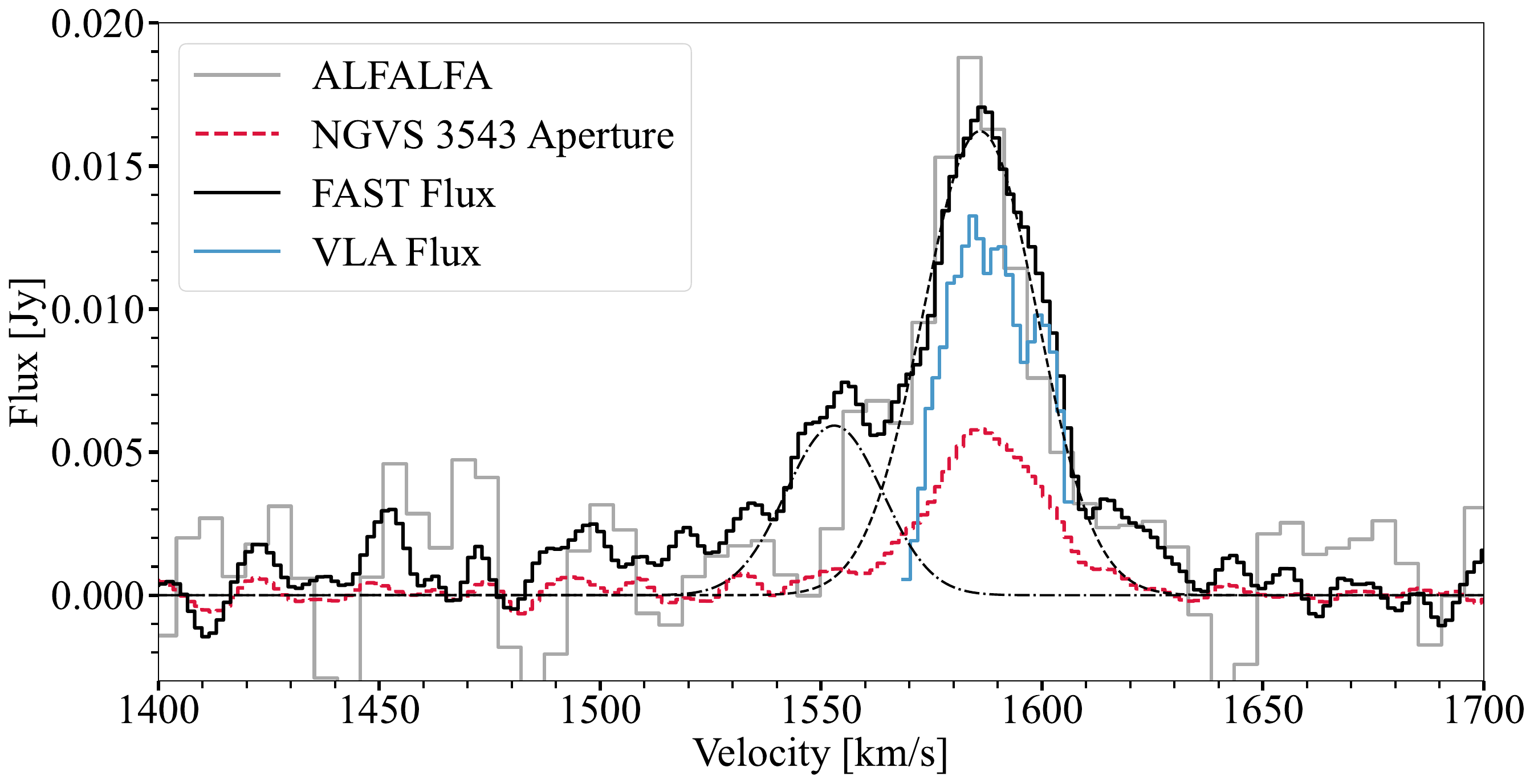}
   \caption{$\hi$ spectra of AGC 226178 from ALFALFA, FAST, and the VLA. The grey curve is the public ALFALFA spectrum from \cite{2018ApJ...861...49Haynes}; the blue curve is for the natural-weighted cube of the VLA; the black curve is the FAST spectrum extracted using a 4.2\arcmin $\times$ 3.5\arcmin elliptical aperture centred on AGC 226178; the dashed line represents the Gaussian fit for the main component, whereas the dash-dotted line corresponds to a fit of a Gaussian to the second part of the system (i.e. the low-velocity extension in the position velocity diagram shown in Fig.\ref{fig: DESI_FAST_PV}). The red dashed curve represents the FAST spectrum extracted from a beam-sized aperture centred on the nearby galaxy NGVS 3543.}
   \label{fig: HI_spectrum}
\end{figure*}

To probe any emission from NGVS 3543 based on the FAST data, we used an aperture of one average beam size in diameter centred on the optical centre of NGVS 3543 (red dash circle in the left panel of Fig.\ref{fig: DESI_FAST_PV}) to extract the $\hi$ spectrum. The spectrum is shown as the red dashed curve in Fig.\ref{fig: HI_spectrum}. Due to the limited spatial resolution of the FAST data, the aperture also includes emission from the main body of AGC 226178.

No significant signal is found near the expected velocity of NGVS 3543. Given the notoriously unreliable nature of projected positions near clusters—particularly in the unrelaxed Virgo cluster—the galaxy’s true velocity could plausibly fall anywhere near or below 3000 $\rm km~s^{-1}$. For this reason, we extended our search to a broader velocity range (500–4000 $\rm km\,s^{-1}$), but still detected no emission, unless NGVS 3543’s velocity were to coincide with that of AGC 226178 and created a blended emission feature. We applied the same aperture-based search to the natural-weighted VLA data cube. No significant detection was found within a beam-sized aperture centred on NGVS 3543. Therefore, we did not find evidence for $\hi$ emission associated with NGVS 3543. Based on the noise level of the data, we place a 3$\sigma$ upper limit of $\sim 3.2 \times 10^5\,M_\odot$ on its $\hi$ mass.

\begin{table*}
\caption{Integrated $\hi$ properties of sources detected in the FAST observation}
\label{tab:fast}      
\centering                          
\begin{tabular}{lccccccc}          
\hline\hline
Galaxy ID & Flux & $V_{\rm cen}$ & $W_{20}$ & $W_{50}$ & S/N & Distance & $\log(M_{\hi})$ \\
          & (Jy\,km\,s$^{-1}$) & (km\,s$^{-1}$) & (km\,s$^{-1}$) & (km\,s$^{-1}$) &  & (Mpc) & ($M_\odot$) \\
\hline
VCC 2037      & $0.67 \pm 0.04$ & 1143.4 & $53.9 \pm 1.6$ & $28.0 \pm 1.2$ & 43.5 & 9.6  & $7.16 \pm 0.10$ \\
VCC 2034      & $1.28 \pm 0.04$ & 1501.1 & $68.3 \pm 1.1$ & $49.3 \pm 0.8$ & 57.2 & 16.5 & $7.92 \pm 0.10$ \\
AGC 226178    & $0.66 \pm 0.04$ & 1580.6 & $66.9 \pm 1.9$ & $27.9 \pm 1.5$ & 36.3 & 16.5 & $7.63 \pm 0.10$ \\
VCC 2015      & $0.35 \pm 0.04$ & 2208.1 & $46.4 \pm 3.8$ & $41.1 \pm 2.9$ & 13.7 & 16.5 & $7.35 \pm 0.10$ \\
\hline
\end{tabular}
\end{table*}

\subsection{Small-scale $\hi$ morphology and kinematics}\label{sec:vlahi}
\begin{figure*}[htp]
   \centering
   \includegraphics[width=18cm]{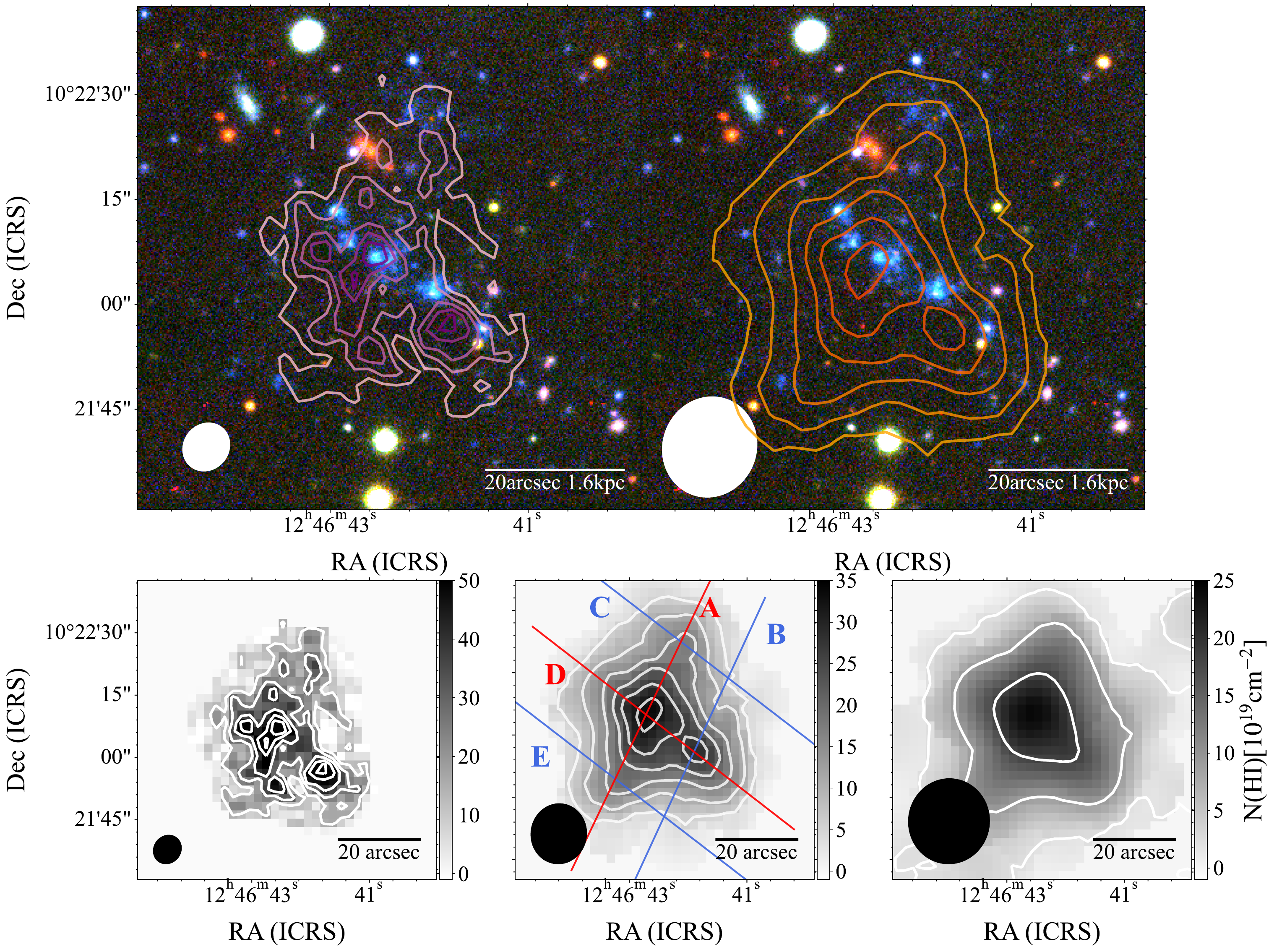}
   \caption{Integrated $\hi$ intensity distribution (moment 0 maps) of AGC 226178. Filled ellipse in the bottom left corner of each panel indicates the synthesized VLA beam of the corresponding data. \textit{Upper left}: $\hi$ intensity contours of the robust-weighted overlaid on the false colour composite image made of NGVS $i$-, $g$-, $u$-band. The contour levels indicate a column density of (1.3, 2.6, 3.9, 5.2, 6.4)$\rm \times10^{20}\,cm^{-2}$. The
scale bar at the bottom right corner is for a distance of 16.5 Mpc; \textit{Upper right}: similar to the upper left panel, but for the natural-weighted $\hi$ map. The contours are for column density of (0.5, 1.1, 1.6, 2.1, 2.7, 3.2) $\rm \times10^{20}\, cm^{-2}$; \textit{Bottom left}: Robust-weighted moment 0 map in grey scale with added contours. Contour levels are the same as the top left panel; \textit{Bottom middle}: Moment 0 map of the natural-weighted map in grey scale and added contours. Contour levels are the same as in the top right panel. The solid lines indicate the slits used to extract the PV diagrams shown in Fig. \ref{fig: VLA_PV_model}. Slit D indicates the major axis with the largest velocity gradient, whereas slit A indicates the minor axis that is perpendicular to the major axis and across the central $\hi$ peak; \textit{Bottom right}: Similar to the other bottom panels, but for the $uv$-tapered map. The contours correspond to column density of (0.2, 0.9, 1.8) $\rm\times10^{20}\,cm^{-2}$.}
   \label{fig: VLA}
   
\end{figure*}

While the FAST observations detect faint and extended gas structures, with poor resolving power on galactic scales, VLA observations can revolve $\hi$ gas in individual objects, albeit at relatively 
worse column density
sensitivity. Fig.\ref{fig: HI_spectrum} compares the integrated $\hi$ spectra of AGC 226178 from ALFALFA, FAST and the VLA. The overall velocity profiles of ALFALFA and FAST are in broad agreement with each other. The VLA observations appear to miss the faint, low-velocity tail shown in the single-dish spectra. Particularly, the natural-weighted image recovers $\rm 0.33\ Jy\ km\,s^{-1}$ in total, which is $53\%$ of the ALFALFA flux. Given the analysis in the previous section, the missing $\hi$ flux of the VLA data is expected to be mostly associated with extended $\hi$ features (resulting from tidal or ram pressure stripping) beyond the main body of AGC 226178.
Fig. \ref{fig: VLA} presents the VLA $\hi$ moment 0 maps produced with robust (top left; bottom left), natural (top right; bottom middle) and $uv$-tapered natural (bottom right) weightings.

Similar to that discovered by \cite{2015AJ....149...72Cannon} based on a D-configuration map, AGC 226178 appears highly centrally concentrated and single-peaked in our lowest resolution map with $uv$-taperred weighting. However, the $\hi$ distribution is resolved into two major overdensities, approximately aligned with the brightest blue stellar clusters along the north-east to south-west direction in the optical main body of AGC 226178. The peak of the central $\hi$ clump has N($\hi$) = $\rm 4.2\times 10^{20}cm^{-2}$, whereas the peak of the clump in the south-west has N($\hi$) = $\rm 3.5\times 10^{20}cm^{-2}$. In addition to the two major peaks aligned along the north-east to south-west direction, the north-east to south-west direction across the central peak is also stretched, matching the spatial locations of the relatively faint blue stellar clusters to the north-west and south-east end \citep[][]{2022ApJ...926L..15Jones}. In the robust-weighted $\hi$ map, which has a linear resolution of $\sim$ 0.6 kpc, the $\hi$ major clump in the natural-weighted map is further resolved into several sub-peaks indicating a hierarchical $\hi$ gas structure in AGC 226178. 

\begin{figure}
\includegraphics[width=9cm]{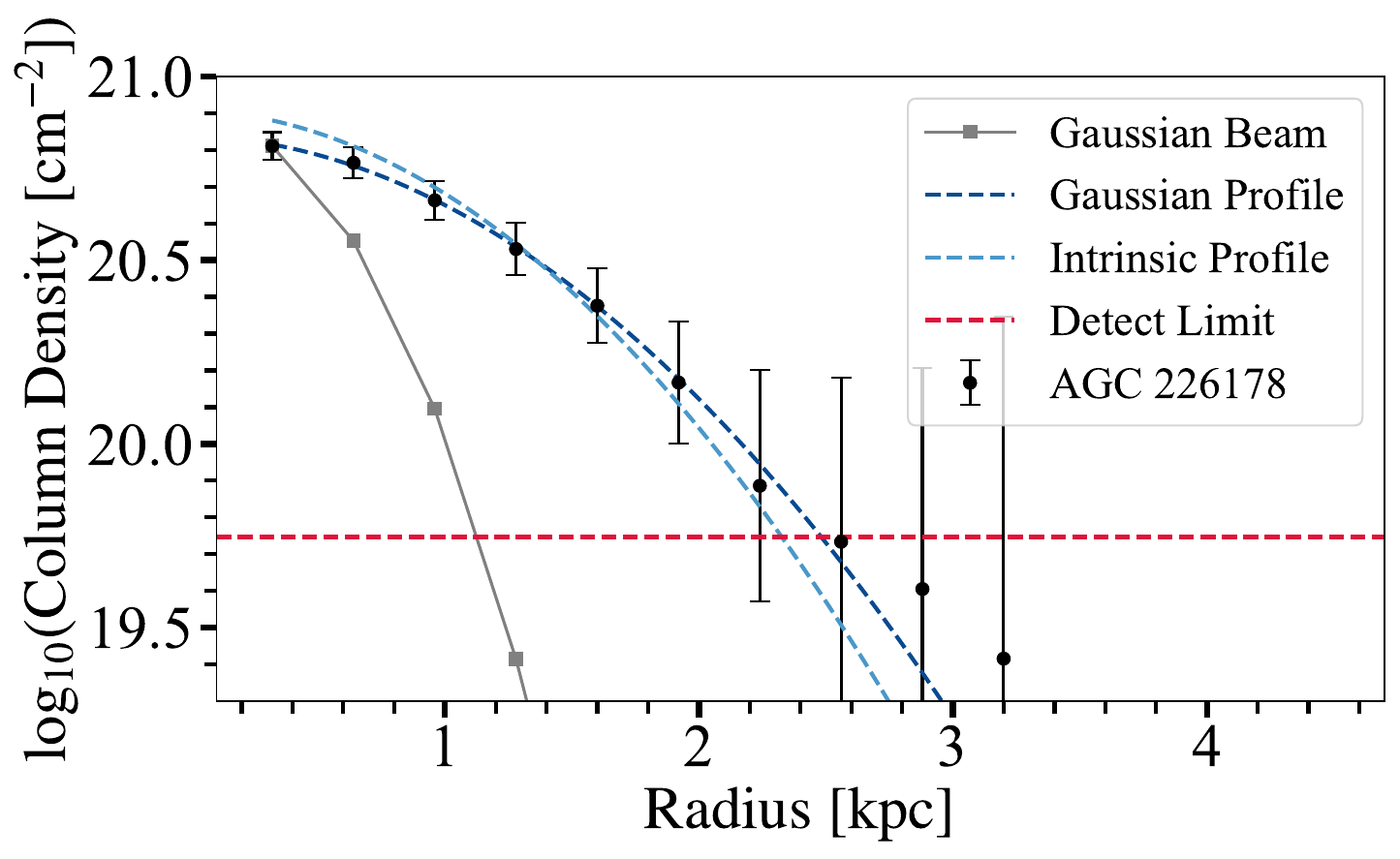}
\caption{The $\hi$ column density radial profile of the natural-weighted VLA image of AGC 226178. The dark blue represents the Gaussian fit to the profile, whereas the light blue represents the deconvolved intrinsic profile. The red dashed line indicates the 3$\sigma$ detection limit, whereas the grey solid line is the beam profile.}
\label{fig: HI_profile}
\end{figure}

\begin{figure*}
   \sidecaption
   \includegraphics[width=12cm]{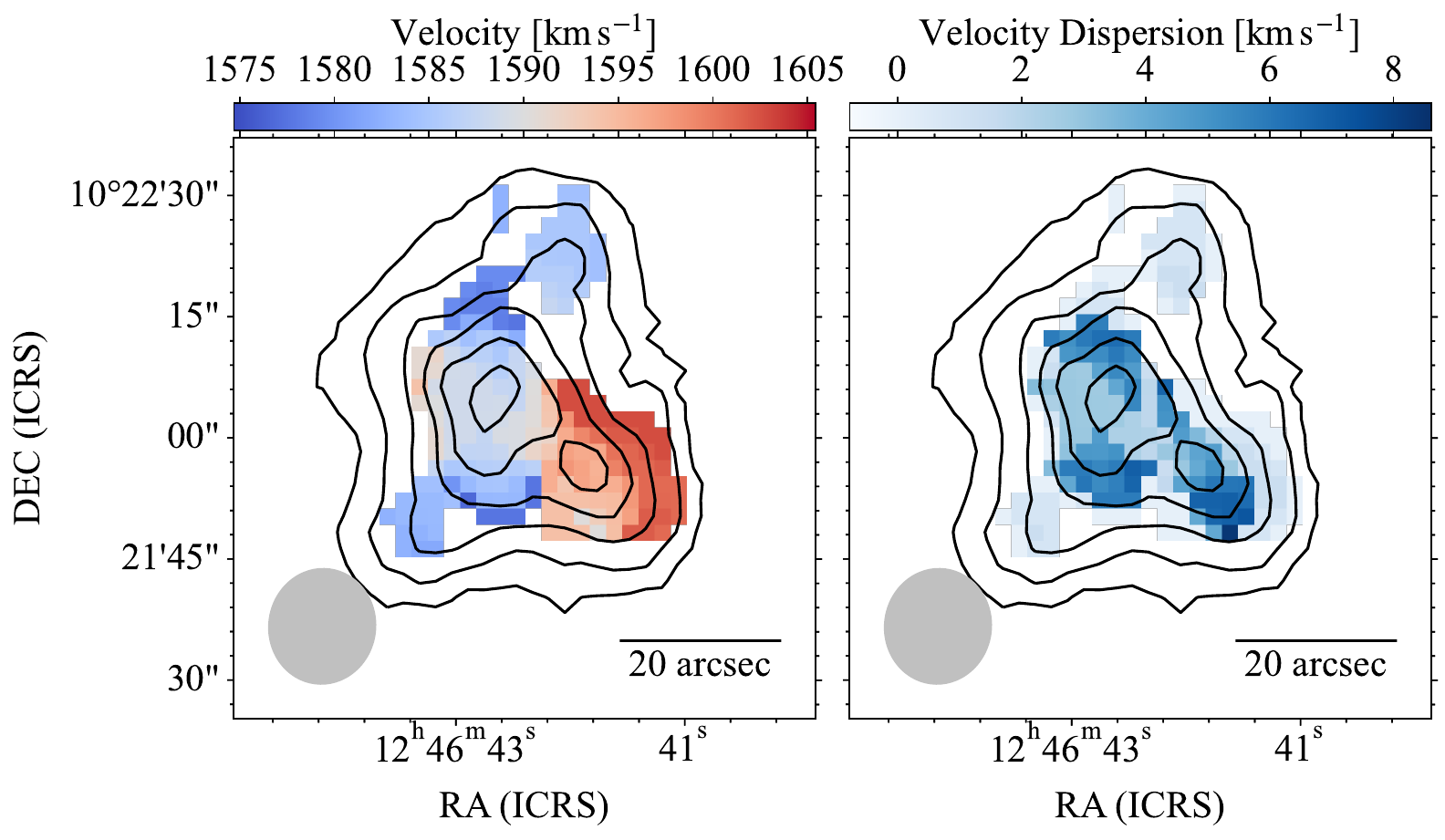}
   \caption{\textit{Left}: Intensity-weighted velocity field (moment 1 map) of the natural weighted VLA data overlaid with the moment 0 map contours shown in Fig. \ref{fig: VLA}; \textit{Right}: Intensity-weighted velocity dispersion map (moment 2 map) of the same VLA data overlaid with moment 0 contours. The synthesized beam is indicated in the bottom left corner of each panel.}
   \label{fig: Mom1}
\end{figure*}

The intensity-weighted velocity field (moment 1) and velocity dispersion map (moment 2) of the natural-weighted image cube are shown in Fig. \ref{fig: Mom1}. Only pixels with signal-to-noise ratio larger than 3 are used to create the two moment maps. A pronounced velocity gradient along the north-east to south-west direction of the stellar main body is evident in the moment 1 map. 
The direction of this velocity gradient matches the elongation of the stellar main body, which may suggest the presence of ordered motion, such as rotation. However, the overall irregular stellar morphology precludes a definitive interpretation.
Notably, this velocity gradient is not evident in the velocity field of the $\hii$ regions \citep{2022ApJ...935...50Bellazzini}.
Based on the spatial distribution in their $\rm H\alpha$ emission line image (see also Fig. \ref{fig: Clumps}), the two main $\hi$ clumps exhibiting velocity gradients are located amidst multiple $\hii$ regions, suggesting that the velocity differences may instead be caused by star formation feedback or supernova explosions. 

The whole cloud exhibits a complex morphology, its optical counterpart appears elongated. It also shows significant extension perpendicular to the main axis. While a velocity gradient is present, its origin is unclear and cannot be conclusively attributed to rotation. The inclination angle is also highly uncertain given the limited spatial resolution. Given the overall ragged yet round-ish morphology of the $\hi$ intensity distribution, we adopt a series of circular apertures to derive the radial profile.
Circularly averaged radial profile of the $\hi$ column density N($\hi$) extracted from the natural-weighted moment 0 map is shown in Fig. \ref{fig: HI_profile}. The measured central N($\hi$) reaches 10$^{20.8}$ cm$^{-2}$ ($\simeq$ 5 $M_{\odot}$ pc$^{-2}$). The N($\hi$) radial profile can be fitted with a Gaussian function. The fitting returns a full width at half maximum (FWHM) of 2.7 kpc. Thus, by subtracting the mean FWHM of the beam in quadrature, the intrinsic FWHM is estimated to be 2.4 kpc. The deconvolved central N$(\hi)$ is $10^{20.9}$ cm$^{-2}$ ($\simeq$ 6 $M_{\odot}$ pc$^{-2}$).

\subsection{$\hi$ velocity field modelling and gravitational binding}\label{sec:hidynmass}

In order to understand the dynamical stability of AGC 226178, we consider two limiting scenarios to interpret the $\hi$ velocity field: one in which the system is partially supported by rotation, with the rotational velocity taken as large as permitted by the $\hi$ observations, and another in which the system is purely pressure-supported. While both scenarios may yield the same integrated velocity width, the required dynamical masses may differ due to their distinct geometries and kinematic structures. For given baryonic mass, the rotation-supported case generally requires lower dynamical mass than does the dispersion-supported case. In the following, we examine these two scenarios in turn.

We modelled the system with a rotational velocity field using the 3D tilted ring fitter $\rm ^{3D}$\verb|Barolo|, which is well-suited for modelling a velocity field that is significantly affected by beam smearing \citep{2015MNRAS.451.3021Di}. Given the complexity of the velocity field of AGC 226178 (Fig. \ref{fig: VLA_PV_model}) and the tentative rotation-like gradient discussed above, we inspected each channel of the natural-weighted image cube and manually masked features that are not associated with the central elongated structure. After the masking step, there remains 0.12 Jy $\rm km\,s^{-1}$, accounting for $31\%$ of the total $\hi$ flux in the natural-weighted cubes and corresponding to a $\hi$ mass of $M_{\rm HI,\  rot}=7.7\times10^6M_{\odot}$.
Although the stellar counterpart of the $\hi$ gas appears highly elongated or inclined (by assuming a disc geometry), considering its extreme nature, we tested different initial values of the inclination angle at 10-degree intervals and allowed it to vary as a free parameter during the fitting. The resultant best-fit inclination is $i=67$ deg. The best-fitting model returns a maximum rotational velocity of $v_{rot}=8.5$ $\rm km\ s^{-1}$. For a well-ordered rotating disc, the velocity width and the maximum rotational velocity follow the relation as $w_{50}\approx 2v_{rot}
$ \citep{2021MNRAS.508.1195Ponomareva}. However, the measured $w_{50}$ from the VLA map (31.7$\pm$3.4~km s$^{-1}$) of AGC 226178 is nearly twice that deduced from the rotational velocity, suggesting that a substantial portion of the observed velocity width arises from non-rotational components such as turbulence or unresolved internal motions within the clumpy gas distribution.

To estimate the dynamical mass of the main body (i.e. excluding the low-velocity tail structure), we adopt the following equation from 
\cite{1996ApJS..105..269Hoffman}:

\begin{equation}
  M_{\rm dyn}=2.325\times 10^5\left(\frac{V^2_{\rm rot}+3\sigma ^2}{\rm km^2~s^{-2}}\right)\left(\frac{r}{\rm kpc}\right)\ M_{\odot},
  \label{eq:dyn}
\end{equation}
where $V_{\rm rot}$ is the rotational velocity of the system, $\sigma$ is the velocity dispersion, in km s$^{-1}$, whereas $r$ is the radius of the system in kiloparsec.

The maximum radius of the area of $\hi$ emission used in the 3D tilted ring fitting is $r$=1.8 kpc. We subtract the best-fit rotation field from the natural-weighted image cube, and then obtain a line of sight velocity dispersion $\sigma$ of 6.9 $\rm km\,s^{-1}$ for the central 1.8 kpc of the rotation-subtracted cube, With the above parameters, $M_{\rm dyn}$ is estimated to be $9\times10^7$ $M_{\odot}$. This estimate of the dynamical mass should be considered a lower limit, as it assumes that the entire velocity gradient arises from ordered rotation. Alternatively, by assuming that the velocity gradient arises from random motions of the clumps and the whole system is governed by velocity dispersion without rotation support, we can obtain an upper limit to the dynamical mass.  
To explore this case, we fit a Gaussian profile to the natural-weighted $\hi$ velocity profile (Fig. \ref{fig: HI_spectrum}) extracted from the central 1.8 kpc, and derive a velocity dispersion $\rm \sigma=17\ km\ s^{-1}$. The resulting maximum dynamical mass is $M_{dyn}=3.7\times10^8M_{\odot}$. 

The total $\hi$ mass obtained by FAST observation is $M_{\rm HI}=4\times 10^7 M_{\odot}$. According to both the spectrum and PV map from FAST, AGC 226178 appears to consist of two distinct components: a main structure and a faint low-velocity tail. To compare with the dynamical mass estimation for the main component of AGC 226178, we performed a two-Gaussian decomposition of the $\hi$ spectrum. The resulting Gaussian profiles are shown in Fig.\ref{fig: HI_spectrum}. Based on the fit, the main component has a flux of 0.53 $\rm mJy\,km\,s^{-1}$, whereas the second component contributes 0.16 $\rm mJy\,km\,s^{-1}$, indicating that the main part contains an $\hi$ mass of $M_{\rm HI_{main}}=3.4\times 10^7 M_{\odot}$.

By multiplying the FAST $\hi$ mass by a factor of 1.36 to account for helium and heavier elements, the total baryonic mass of the main part of AGC 226178 is estimated to be $M_{\rm bary,\ main} \approx 4.6 \times 10^7\, M_{\odot}$, with negligible stellar contribution. The resulting dynamical-to-baryonic mass ratio is $M_{\rm dyn}/M_{\rm bary} \simeq 2.0$ if rotational support is assumed, and up to $8.0$ in the case of purely dispersion support. Even in this latter case, the ratio remains relatively modest compared to typical values for dark matter dominated galaxies \citep{2022NatAs...6...35Lelli}, suggesting that a strongly dark-matter-dominated nature for AGC 226178 is unlikely.

If the system is not gravitationally bound, the concept of dynamical mass becomes physically ambiguous and loses its interpretative validity. Nevertheless,
the $M_{\rm dyn}/M_{\rm bary}$ ratio is equivalent to the well-known virial parameter \citep[][]{Bertoldi1992}, differing only in the pre-factor, which has been used to determine whether a system is gravitationally bound or not. 
The excess $M_{\rm dyn}/M_{\rm bary}$ ratio of AGC 226178 suggests that the system is either gravitationally unbound or requires additional support to remain stable, potentially from external pressure or an unobserved mass component in the system.

\subsection{Resolved $\hi$ clumps in AGC 226178}\label{sec: hiclumps}

\begin{figure*}[htp]
   \centering
   \includegraphics[width=18cm]{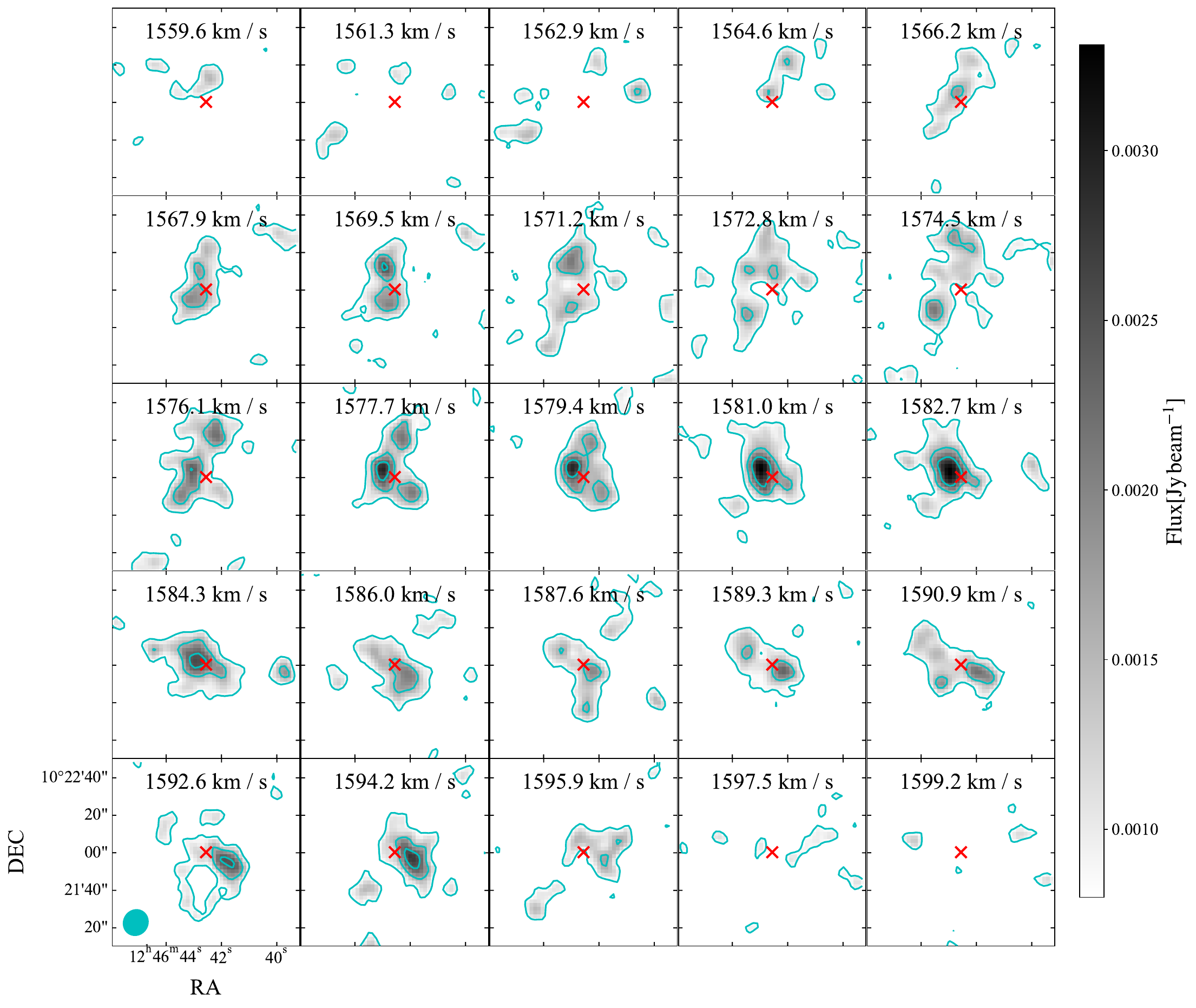}
   \caption{Natural-weighted channel maps. The contours represent 2,4,and 6 times the RMS noise of 0.4 $\rm mJy\,beam^{-1}\,channel^{-1}$. The red cross indicates the centre of AGC 226178, at (RA, Dec)=$\rm 12^h46^m42^s.56,\ 10^o22'00''.32$}
   \label{fig: Channel_map}
\end{figure*}

The VLA observations reveal that the $\hi$ distribution in AGC 226178 is highly fragmented and clumpy. To quantify the properties of the $\hi$ clumps, we used the \verb|CLUMPFIND| algorithm in the Starlink package \verb|CUPID|\footnote{https://starlink.eao.hawaii.edu/starlink/CUPID} to detect clumps on the naturally weighted data cube. 
Although the higher-resolution robust-weighted cube appears to reveal more structures, only the relatively bright ones can be robustly detected and measured, due to the poorer sensitivity. So here we focus on the natural-weighted image cube. As input for running \verb|CLUMPFIND|, an RMS noise level of 0.4 $\rm mJy\,beam^{-1}$, measured from the natural-weighted cube, is used. Pixels below 1 $\times$ RMS are excluded from detection, each clump must have a spatial area greater than the beam area, and adjacent peaks are considered disjoint only if the dip between them exceeds 0.2 $\times$ RMS and their separation is greater than 2 pixels. To illustrate the clumpy $\hi$ structures, the channel maps are presented in Fig. \ref{fig: Channel_map}.

\begin{figure*}[htp]
    \centering
    \includegraphics[width=18cm]{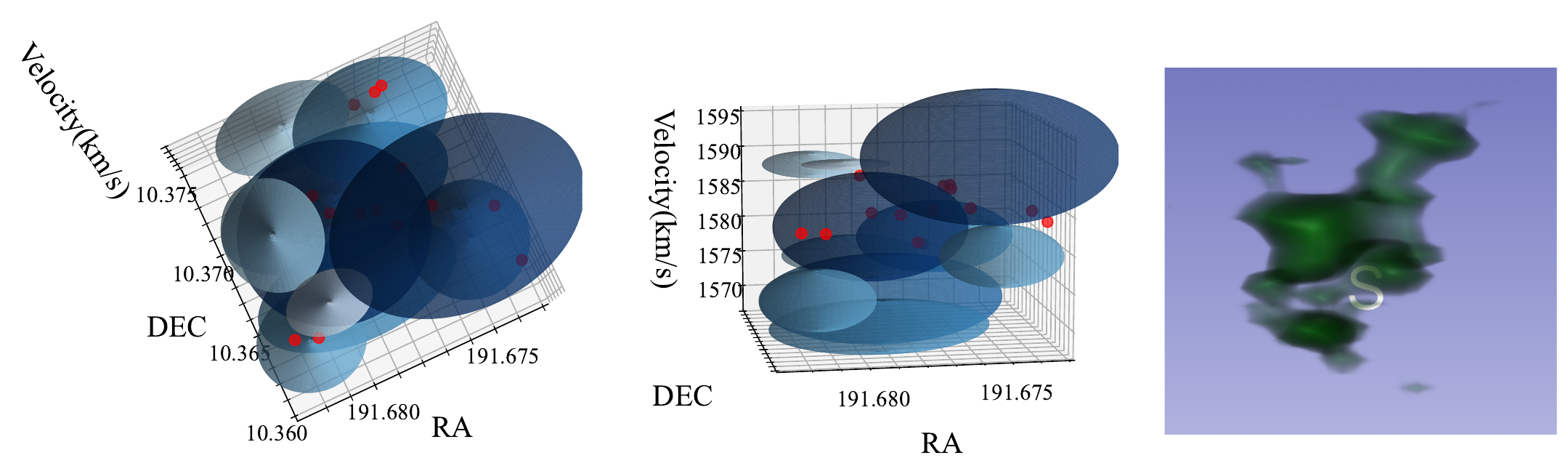}
    \label{fig: CUPID_test}
    \caption{\textit{Left and middle}: 3D illustration of $\hi$ clumps detected with CUPID, with different viewing angles for the {\it left} and {\it middle} panels. Blue ellipsoids represent the $\hi$ clumps, and red dots indicate the H$\alpha$ emission regions detected in MUSE data. The size of the ellipsoids indicates the spatial size and velocity width of each clump, colour-coded by the $\hi$ flux. \textit{Right}: 3D illustration of the natural-weighed VLA data produced by the 3D-Slicer package, with a fixed threshold cut applied and the same viewing angle as the middle panel.}
   \label{fig: CUPID}
\end{figure*}

\begin{figure*}[htp]
   \centering
   \includegraphics[width=18cm]{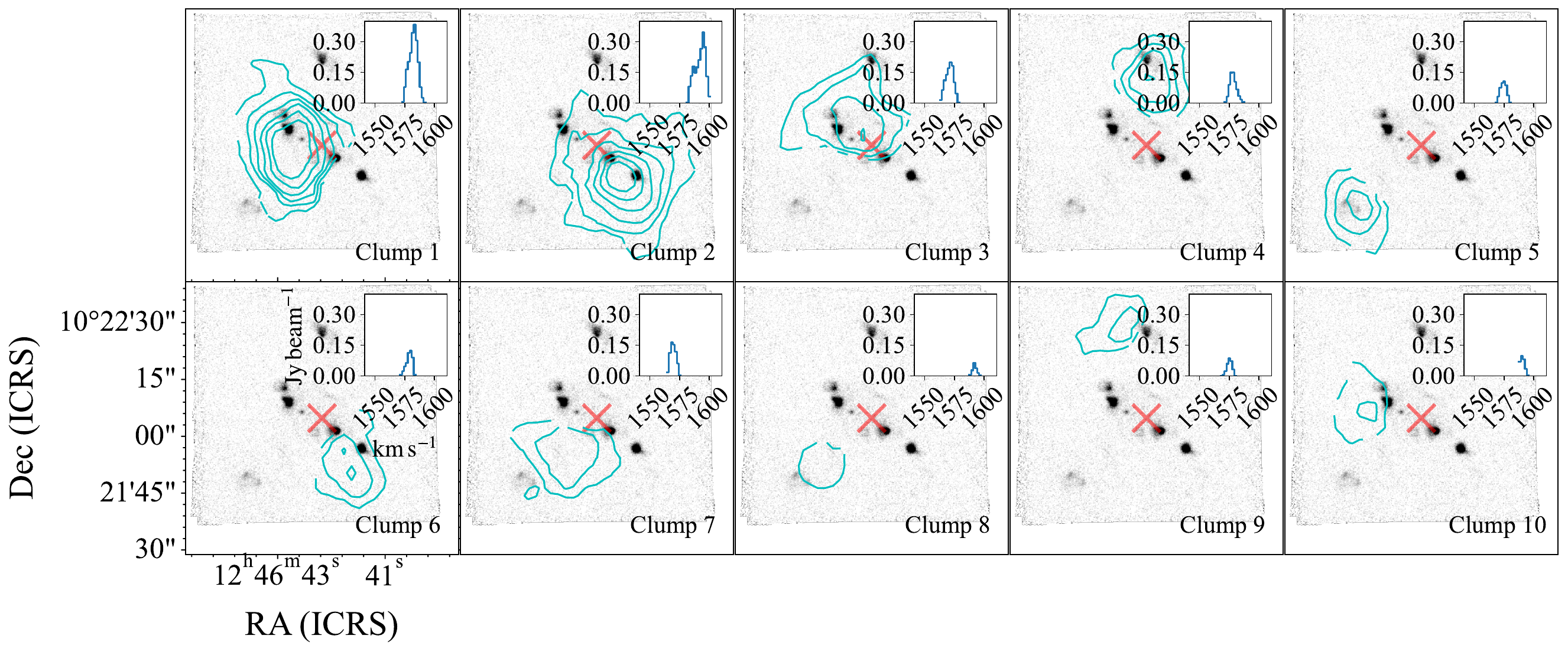}
   \caption{Contour maps of the brightest $\hi$ clumps detected in the natural-weighted VLA image cube. The background displays the H$\alpha$ line emission image constructed from the MUSE datacube. The inset plot in each panel shows the $\hi$ spectrum of each clump. The red cross indicates the centre of AGC 226178.}
   \label{fig: Clumps}
\end{figure*}

A total of 23 clumps were detected by \verb|CUPID|. After visual inspection, we only kept the brightest 10 clumps which have an integrated signal-to-noise ratio larger than  $8\sigma$. The relevant properties of these clumps are listed in Table \ref{tab:clumps}. The reported spatial size of each clump has been corrected for beam-smearing by deconvolving the instrumental beam width, assuming a Gaussian profile, and the velocity width has been similarly corrected for the velocity resolution of the data. Distribution of the clumps in the 3D space of RA, DEC, and velocity is illustrated in Fig. \ref{fig: CUPID}, where the left and middle panels are for two different viewing angles. Each clump is plotted as an ellipsoid, of which the length of the three principle axes is, proportional to the spatial size and velocity width, and the ellipsoid is colour-coded according to the integrated flux of each clump. The right panel shows a 3D visualization of the image cube created with \verb|Slicer|.\footnote{https://www.slicer.org/} We also present the $\hi$ intensity map and velocity profiles of these clumps in Fig. \ref{fig: Clumps}.

As is clearly illustrated in Fig. \ref{fig: CUPID}, the two brightest clumps in AGC 226178 have a systemic velocity difference that dominates the rotation-like velocity gradient mentioned in previous sections. The remaining, fainter clumps mostly occupy the low-velocity end, echoing the more pronounced stripping signature at the lower-velocity end revealed by the FAST data.

We point out that, as with previous similar studies in the literature, the search for $\hi$ clumps is strongly limited by the resolving power of the data. This is because the structure of the interstellar medium is hierarchical in nature, as demonstrated by the finer details revealed in our robust-weighted maps (e.g. Fig. \ref{fig: VLA}). Moreover, the size of the clumps (Table \ref{tab:clumps}) reaches up to kiloparsec scales which is an order of magnitude larger than the largest gravitationally bound giant molecular clouds. Therefore, the $\hi$ clumps identified here represent the finest overdensities in the $\hi$ distribution that can be resolved by our VLA data.

\subsection{Relation between $\hi$ gas and star formation}
\label{sec:sf}

We use the H$\alpha$ emission from star-forming regions as a tracer of the current star formation in AGC 226178. AGC 226178 has previously been observed using Very Large Telescope (VLT)/Multi Unit Spectroscopic Explorer \citep[MUSE,][]{2014Msngr.157...13Bacon}, as part of the observing programme 0101.B-0376A (P.I: R. Mu\~noz), the results have been presented in \cite{2022ApJ...935...50Bellazzini}. The MUSE 1\arcmin $\times$ 1\arcmin~field for AGC 226178 (also known as BC3 in \cite{2022ApJ...935...50Bellazzini}), was centred at RA, DEC = (191.677299, 10.36919), observed on 28 Feb 2019, with a seeing disc $\rm FWHM \simeq 1.^{\prime\prime}1$. 
$\hii$ regions were identified by \citet{2022ApJ...935...50Bellazzini} using a narrow-band image created by integrating a 5$\rm \AA$ spectral window centred on the H$\alpha$ emission line. Regions with peak intensities exceeding 3$\sigma$ above the background were selected and H$\alpha$ fluxes were measured within circular apertures of radius $1\farcs5$. We make direct use of the regions identified and characterized in their analysis; the details of the identification and measurement procedures can be found in their paper.
The greyscale background in Fig.~\ref{fig: Clumps} displays continuum-subtracted H$\alpha$ line emission image from MUSE, integrated over the wavelength range 6591.54–6605.29\text{\AA}. The plot illustrates the projected spatial relationship between the $\hii$ regions and $\hi$ clumps.

We associate the $\hi$ clumps identified in Section \ref{sec: hiclumps} with the $\hii$ regions based on their proximity in both spatial location and velocity. It turns out that 4 $\hi$ clumps (ID: 1, 2, 4, 5) have associated $\hii$ regions, which are indicated in Table \ref{tab:clumps}.
The star formation rate (SFR) for the whole system and each individual $\hii$ region was calculated from the $\rm H\alpha$ line flux (corrected for extinction using the Balmer decrement technique) of the $\hii$ regions, employing the calibration provided by \cite{1998ApJ...498..541Kennicutt}, which assumes a Salpeter initial mass function (IMF). However, we note that this $\rm H\alpha$–SFR conversion may be uncertain in our case, as the assumption of constant (stationary) star formation over timescales of several Myr may not hold. Moreover, at such low SFR levels (especially for $\log(\rm SFR / M_{\odot},yr^{-1}) \lesssim -3$), stochastic sampling of the IMF may lead to an underestimation of the total star formation \citep[see e.g.][]{2009ApJ...706.1527Boselli, 2018A&A...615A.114Boselli}.

\begin{figure*}[htp]
   \centering
   \includegraphics[width=16cm]{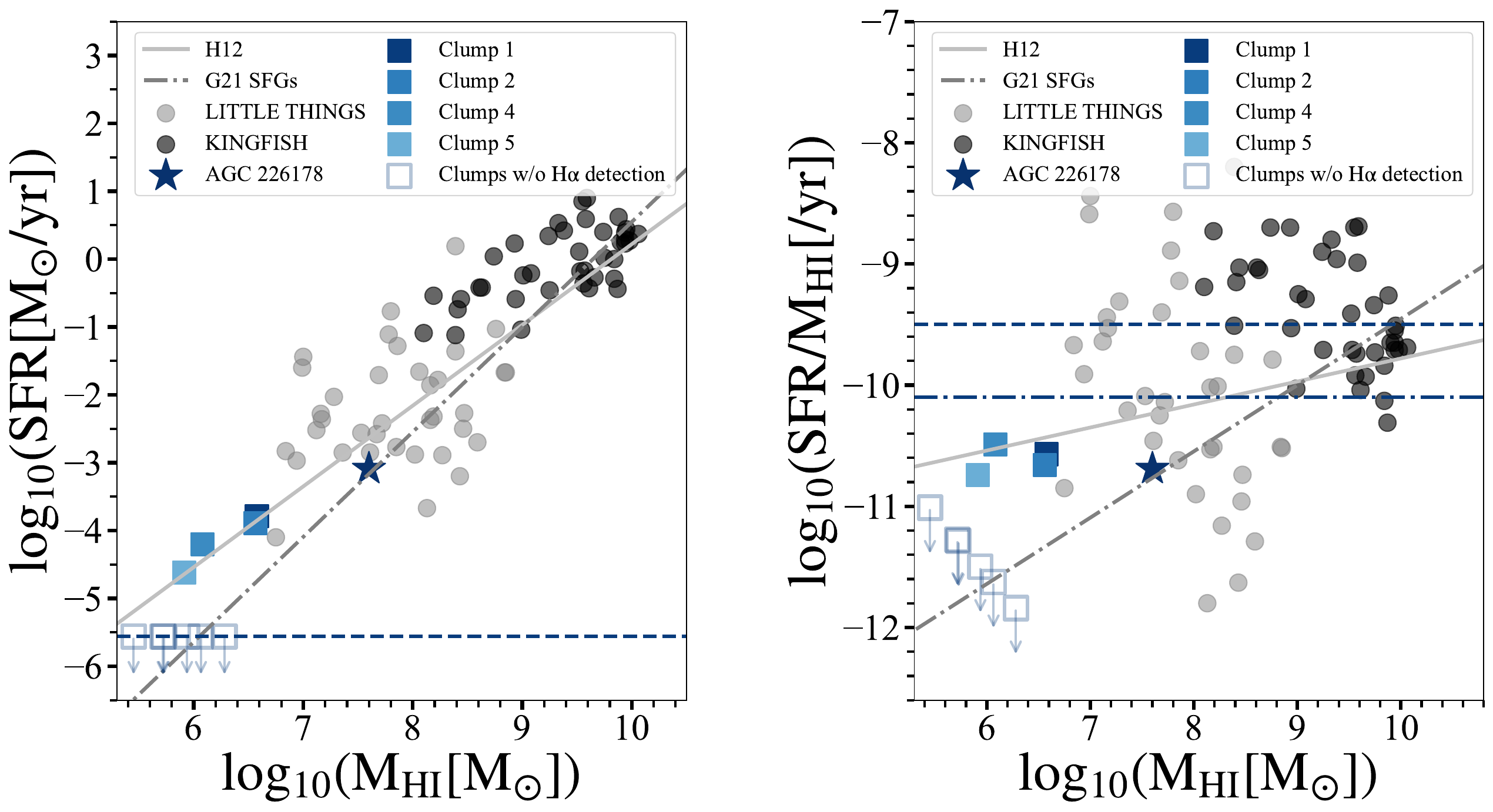}
   \caption{\textit{Left}: $\hi$ mass vs the star formation rate (SFR) diagram for the $\hi$ clumps, AGC 226178 as a whole and literature samples of nearby galaxies;   \textit{Right}: $\hi$-based star formation efficiency vs $\hi$ mass. The solid blue squares represent the $\hi$ clumps associated with $\hii$ regions in AGC 226178, whereas the hollow squares represent $\hi$ clumps not associated with star formation. The SFR for the hollow squares are arbitrarily set to the 3$\sigma$ detection limit of SFR. The grey and black circles are star-forming galaxies from the Local Irregulars That Trace Luminosity Extremes, The HI Nearby Galaxy Survey (LITTLE THINGS) sample \citep{2012AJ....144..134H} and Key Insights on Nearby Galaxies: a Far-Infrared Survey with Herschel (KINGFISH) sample \citep{2011PASP..123.1347Kennicutt}. The grey solid and dash-dotted lines indicate the observed average relation of nearby galaxies from \cite{2012ApJ...756..113H} and \cite{2021ApJ...918...53Guo}, respectively. In the right panel, the dashed horizontal line corresponds to the reciprocal of the Hubble time, whereas the dash-dotted horizontal line indicates the average value obtained by \cite{2010MNRAS.408..919Schiminovich} for the GASS sample.}
   \label{fig: SFR}
\end{figure*}

In Fig. \ref{fig: SFR}, we explore the relation between the SFR and mass of the $\hi$ clumps. Literature samples of nearby galaxies and AGC 226178 as a whole are also plotted for comparison. There is a correlation between the $\hi$ mass and SFR, but with substantial scatter. The $\hi$ clumps associated with star formation (blue squares in Fig. \ref{fig: SFR}) in AGC 226178 generally follow the relation observed for nearby galaxies (left panel). The $\hi$-based star formation efficiency SFR/$M_{\rm HI}$ of AGC 226178 and its clumps fall at the lower end of the range spanned by nearby galaxies (right panel). This broad consistency of AGC 226178 with normal galaxies supports the view that the $\hi$-SFR relation does not reflect a causal connection, but rather is a manifestation of star formation being self-regulated by heating (partly contributed by star formation) and radiative cooling of the interstellar medium \citep[e.g.][]{2005ApJ...634.1067Taylor}.

\begin{figure*}[htp]
   \centering
   \includegraphics[width=16cm]{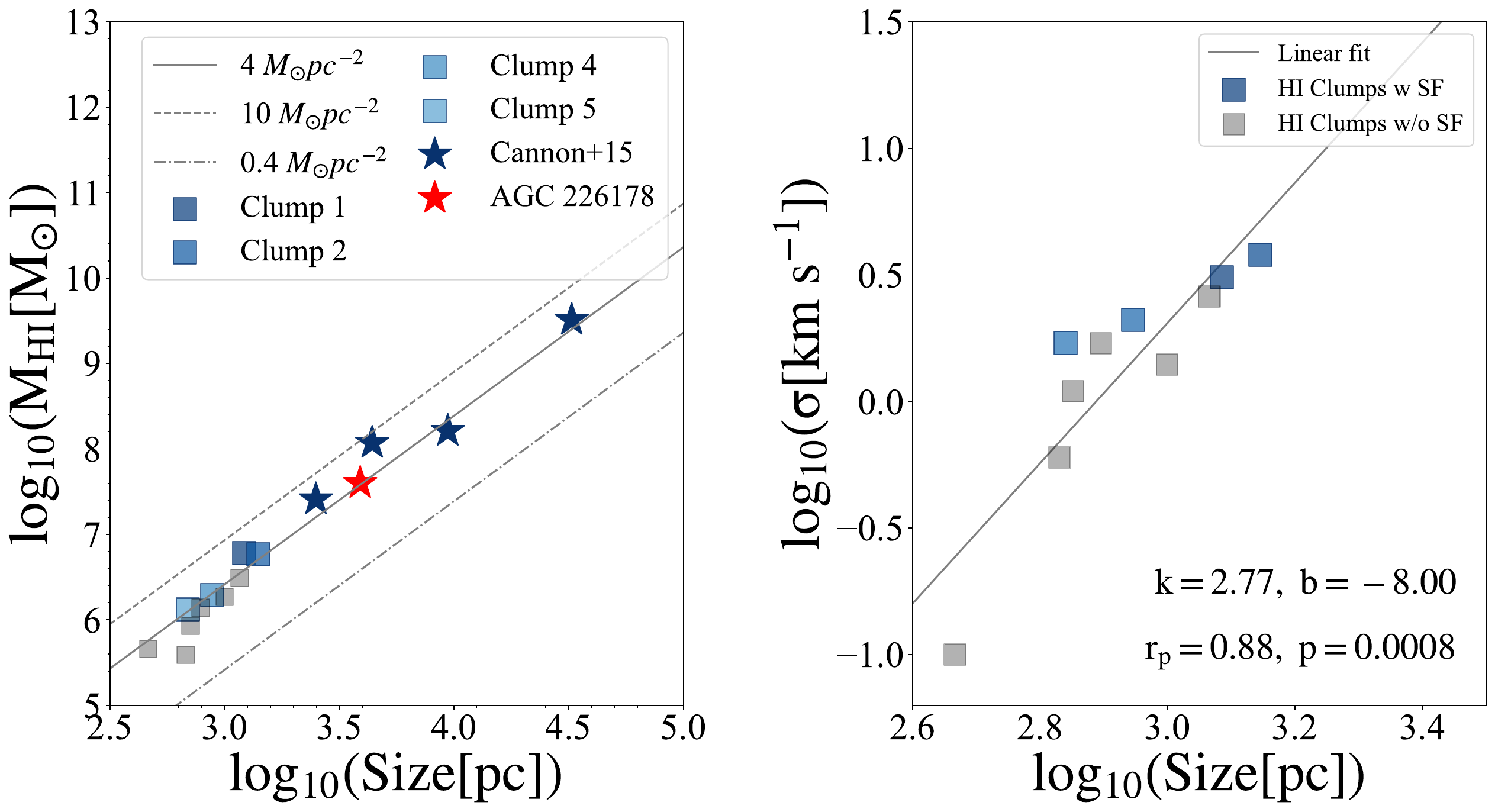}
   \caption{\textit{Left}: The $\hi$ mass-size relation of $\hi$ clumps in AGC 226178 and literature sample of almost dark $\hi$ clouds. The size of each clump is computed as $D = 2\sqrt{A/\pi}$, where $A$ is the projected area estimated from the geometric mean of the major and minor axes in Table. \ref{tab:clumps}.  Dashed lines indicate the critical surface density for $\hi$-to-$\rm H_2$ transition predicted by different models and physical conditions, as indicated in the legend. Stars indicate five almost dark $\hi$ clouds investigated by \cite{2015AJ....149...72Cannon}. The grey lines mark the critical surface density for the atomic-to-molecular gas transition, as predicted by some models. For a detailed discussion of the physical significance of different thresholds, see Section~\ref{sec:sf}. \textit{Right}: the size versus velocity dispersion distribution of $\hi$ clumps. In both panels, $\hi$ clumps associated with $\hii$ regions are plotted with blue filled squares, whereas the others are plotted with filled grey squares. The solid line in the right panel represents the best-fit size-velocity dispersion relation. The best-fit logarithmic slope ($k$), intercept ($b$), and the Pearson's correlation coefficient and p-value are also indicated in the right panel.}
   \label{fig: Scale}
\end{figure*}

In Fig. \ref{fig: Scale}, we further explore the structural properties of the $\hi$ clumps and their connection to star formation. The $\hi$ mass is plotted against clump size in the left panel, and the velocity dispersion of the clumps is plotted against size in the right panel. Mass-size relations corresponding to several characteristic surface densities are also plotted for comparison. Particularly, an $\hi$ surface density of $\rm 10 \ M_{\odot}~pc^{-2}$ is about the observed maximum on (sub)kiloparsec scales of nearby galaxies \citep[e.g.][]{Wong2002}. It is thought that at higher surface densities $\hi$ is mostly transformed to molecular hydrogen. A surface density of $\rm 4 \ M_{\odot}~pc^{-2}$ is about the theoretical critical value for a gas cloud (in protogalactic discs) that can be self-shielded from cosmic background radiation \citep[][]{2005ApJ...634.1067Taylor}.

In the left panel of Fig. \ref{fig: Scale}, we also overplot the five almost dark $\hi$ clouds (including AGC 226178) from \cite{2015AJ....149...72Cannon}. There is a tight $\hi$ mass-size correlation for these $\hi$ sources, similar to the well-known tight correlation observed for ordinary galaxies \citep[e.g.][]{Broeils1997,Wang2016}. The sense of the correlation is broadly consistent with nearly constant average surface densities in different systems. The existence of a similarly tight mass-size correlation for $\hi$ clumps and almost dark $\hi$ clouds suggests a self-regulated structural property of interstellar neutral $\hi$ gas, regardless of environment. With that being said, we note that all of the $\hi$ sources plotted in Fig. \ref{fig: Scale} lie below the line of $\rm 10 \ M_{\odot}~pc^{-2}$ maximum surface density mentioned above, suggesting that these clouds or clumps are $\hi$-dominated. In addition, the star formation-free $\hi$ clumps have systematically lower surface densities than the clumps with star formation and other almost dark clouds. The average surface densities of the star formation-free clumps are systematically lower than the theoretical $\rm \Sigma_{crit}=4\ M_{\odot}~pc^{-2}$ mentioned above.   The intracluster radiation field (ignored in the theoretical calculation) is generally much stronger than the cosmic background radiation field.
 It is therefore quite possible that the $\hi$ clumps without associated star formation are not sufficiently self-shielded, which may inhibit the formation of molecular hydrogen and, consequently, suppress star formation.

In the $\hi$ clump size-velocity dispersion plot shown in the right panel of Fig. \ref{fig: Scale}, there is a significant correlation (the Pearson correlation coefficient $r_{p}$ = 0.86), which is expected for a hierarchical fragmentation governed by turbulence. The best-fit relation is
\begin{equation}
  \log(\sigma~[\rm km~s^{-1}]) = 2.77 \times\log(D~[pc]) -8.00.
  \label{eq:velocity}
\end{equation}
The logarithmic slope of 2.77 is much steeper than that \citep[$\simeq$ 0.4-0.5; e.g.][]{1981MNRAS.194..809Larson, Heyer2009} of the Larson relations for molecular clouds. A cloud dominated by self-gravity has a logarithmic slope close to 0.5, by assuming a constant surface density. Therefore, our $\hi$ clumps are not expected to be bound by gravity. In addition, we note that the velocity dispersion of clumps associated with star formation is systematically (for a given size) larger than those not associated with star formation. This suggests a significant feedback of star formation on the neighbouring $\hi$ gas.

\section{Discussion}
\subsection{The connection between AGC 226178 and nearby galaxies}

Three galaxies (VCC 2037, VCC 2034 and NGVS 3543) with small angular separation from AGC 226178 have been considered in the literature as candidates for having a physical connection with AGC 226178. As pointed out by \cite{2022ApJ...926L..15Jones}, the most probable distance to AGC 226178 is close to the average of the Virgo cluster (16.5 Mpc), given its velocity and resolved stellar colour-magnitude diagram. VCC 2037 has a Virgo-like velocity, and as shown in the Fig.\ref{fig: vla_field}, VCC 2037 exhibits a significantly disturbed structure. However, its distance has been estimated to be 9.6 Mpc based on the Tip of Red Giant Branch (TRGB) measurements, and it is therefore unlikely to be the host galaxy of AGC 226178.
NGVS 3543 has no radial velocity measurement, but its preferred distance is much closer than the Virgo cluster, according to its resolved stellar colour-magnitude diagram \citep[][]{2022ApJ...926L..15Jones}. Moreover, the metallicity of NGVS 3543 is expected (given its stellar mass) to be an order of magnitude lower than that measured in AGC 226178. Therefore, it is unlikely that AGC 226178 originated from NGVS 3543. The distance of VCC 2034 is uncertain, but it is likely to be in the Virgo cluster, given its similar velocity to AGC 226178. 

\cite{2022ApJ...926L..15Jones} uncovered a faint, low signal-to-noise $\hi$ bridge between AGC 226178 and VCC 2034, based on the ALFALFA observation. If the $\hi$ bridge were confirmed, it would be a strong evidence for a physical connection between the two systems. Based on this discovery, \cite{2022ApJ...926L..15Jones} came to the conclusion that AGC 226178 was formed from a gas stripped from VCC 2034. 
The unusually high gas-phase metallicity ($\sim$ 0.5 $Z_{\odot}$) of AGC 226178 indicates that its gas was probably from a galaxy with a much higher stellar mass \citep[$\sim$ 10$^{8.4}$ $M_{\odot}$;][]{2022ApJ...926L..15Jones} rather than self-enriched. The model from  \cite{2022A&A...667A..76Junais} showed that VCC 2034 may have experienced a ram-pressure stripping event starting $\sim$ 150 Myr ago, losing about $3\times 10^8 M_{\odot}$ of gas, which could comfortably account for the formation of AGC 226178.

However, based on our deep FAST observation, we do not find any bridge-like $\hi$ structure connecting AGC 226178 and VCC 2034. The ALFALFA bridge-like feature shows up merely as a 1$\sigma$ extension in the position-velocity space in Fig. 3 of \cite{2022ApJ...926L..15Jones}. That said, we detect a short extension at the lower-velocity end of AGC 226178 that points towards VCC 2034 (Fig. \ref{fig: DESI_FAST_PV}), but VCC 2034 exhibits an asymmetric velocity feature (likely due to ram pressure stripping) extending to the lower-velocity end along the line-of-sight direction, rather than the direction to AGC 226178. Moreover, our VLA data reveal a cometary-like $\hi$ spatial distribution for VCC 2034, with the elongation more closely aligned with the north-south direction than with the direction of the tail of AGC 226178. Above all, it is unlikely that AGC 226178 originated as gas clouds stripped from VCC 2034. 

Besides the neutral $\hi$ gas, ionized gas emission may also be used to probe the environmental stripping event \citep[][]{2022ApJ...935...51J}. According to the radiative hydrodynamical simulations of \cite{2022ApJ...928..144Lee}, besides the relatively bright H$\alpha$ emitting star-forming regions, there exists diffuse ionized gas in the ram pressure stripped tails. Diffuse H$\alpha$ emission from the stripped tails may be detected at surface brightness levels fainter than 6$\times$ 10$^{38}$ erg s$^{-1}$ kpc$^{-2}$. This is within the reach of the Virgo Environmental Survey
Tracing Ionised Gas Emission (VESTIGE) survey \citep[][]{Boselli2018}, whose 1$\sigma$ surface brightness sensitivity is 3.4$\times$ 10$^{38}$ erg s$^{-1}$ kpc$^{-2}$ for a distance of 16.5 Mpc \citep[][]{2022ApJ...935...51J}. 
However, as discussed in detail by \citet{2022A&ARv..30....3Boselli}, this ionized phase may be short-lived. In RPS-affected galaxies, gas is likely stripped as neutral hydrogen and then becomes ionized and eventually heated when mixing with the intracluster medium (ICM). Consequently, diffuse ionized gas is only observed in a handful of late-type galaxies in Virgo, most of which are currently undergoing active ram pressure stripping \citep{2020A&A...644A.161Longobardi,2022A&A...659A..45Sardaneta}. Nevertheless, given the signs of potential environmental interaction in AGC 226178, we examined
the VESTIGE narrow-band H$\alpha$ image, we did not find any diffuse emission in between and beyond the optical main body of AGC 226178 and VCC 2034 at the current epoch.

AGC 226178 thus appears more likely to be a free-floating, isolated system at the present time. However, this does not rule out the possibility that AGC 226178 resulted from a past tidal or ram pressure stripping event and it was completely detached from its parent galaxies and has since evolved independently in the cluster environment, leaving us without an obvious point of origin.   

\subsection{The nature of AGC 226178: A stripped disintegrating gas cloud vs a mini-halo}

Section \ref{sec:hidynmass} suggests that the true dynamical mass of AGC 226178 likely lies between two and eight times its currently observed baryonic mass.
This means that either AGC 226178 is not gravitationally bound and will disintegrate in the near future or that a substantial amount of invisible baryonic or non-baryonic matter exists within it.

\subsubsection{A possible former dark galaxy embedded in mini-halo}

Considering the potential contribution from non-baryonic matter,
here we briefly discuss the possibility that AGC 226178 was a genuine dark galaxy embedded within a mini dark matter halo and its star formation was triggered in the past few hundred million years after falling into the Virgo cluster. In the mini-halo scenario, the significant metallicity of AGC 226178 could be attained through mixing with the hot ICM (e.g. \citealt{2021ApJ...911...68Tonnesen}).

According to the standard $\Lambda$CDM paradigm, luminous galaxies can only form in the centre of dark matter halos above a certain redshift- and environment-dependent threshold mass. Below the threshold mass, the gas cannot cool efficiently to form stars and, in the least massive halos, even the gas component may evaporate completely due to the shallow potential well of the halo \citep{1992MNRAS.256P..43Efstathiou,1996ApJ...465..608Thoul,2024ApJ...962..129L}. 

Two populations of dark galaxies (without star formation) were explored by \cite{2013ApJ...763L..41B} and 
\cite{2017MNRAS.465.3913Bentez_Llambay} with cosmological hydrodynamical simulations. Specifically, the two populations are denoted   REionization-Limited $\hi$ Clouds (RELHICs) and COSmic WEB Stripped systems (COSWEBs). RELHICs are minihaloes that inhabit low-density environment and remain below the star formation threshold but are massive enough to retain baryons against photoevaporation by the cosmic UV background. COSWEBs are minihaloes that are in high-density environment and have been deprived of all baryons due to ram pressure stripping of the cosmic web. The halo mass range of these dark galaxies is $ 10^6 \lesssim M_{200}/M_{\odot}h^{-1} \lesssim 5\times 10^9$ and the baryonic mass $\lesssim$ 10$^{8}$ $M_{\odot}$ \citep[][]{2017MNRAS.465.3913Bentez_Llambay}. Obviously, RELHICs set an upper limit for the halo mass and gas mass of these two populations of dark galaxies. 

The gas component in RELHICs is approximately in hydrostatic and thermal equilibrium, with a thermal velocity dispersion and a nearly spherical shape. The neutral $\hi$ gas is expected to occupy the very central region of the dark haloes, accounting for a very small fraction of the gaseous content. The central $\hi$ surface density ($\sim$ 4$\times$10$^{20}$ cm$^{-2}$) of AGC 226178 matches the upper limit predicted for RELHICs. Nevertheless, in addition to its irregular shape and a velocity broadening likely dominated by turbulence, AGC 226178 has an $\hi$ gas mass (4$\times$10$^{7}$ $M_{\odot}$, for a distance of 16.5 Mpc) that is an order of magnitude larger than the upper limit predicted for RELHICs. The $\hi$ mass would match the predicted upper limit if AGC 226178 were at a distance of $\sim$ 5 Mpc, in the foreground of Virgo cluster. But this would be inconsistent with the notable environmental stripping features in AGC 226178, such as the highly asymmetric $\hi$ velocity profile and gas stripping tails. Taken together, these properties suggest that AGC 226178 is unlikely to be consistent with the characteristics of a RELHIC- or COSWEB-type system. Instead, the observed features may be more plausibly explained by an alternative evolutionary scenario, such as environmental processing through ram pressure stripping or tidal interactions within the Virgo cluster.

\subsubsection{The survivability in the Virgo cluster: A test of ram pressure stripping}\label{sec:rps_test}

Assuming that this system was stripped from another galaxy and has since remained as a free-floating cloud in the cluster for a significant period of time—meanwhile experiencing ram pressure—the relatively high dynamical mass could be attributed either to its dynamical instability or to the presence of unseen baryonic matter. For example, \citet{2007Sci...316.1166Bournaud} found that tidal dwarf galaxies can exhibit dynamical masses approximately twice their visible gas mass, likely due to CO-dark molecular gas.

To better understand the situation in our case, assuming that $\hi$ gas resides in a disk-like structure in order to maximize the gravitational anchoring force,
we model the 3D gravitational potential of the observed galaxy, and test its robustness to ram pressure. We use a triple Miyamoto-Nagai model \citep{2015MNRAS.448.2934Smith}. In this model, three separate Miyamoto-Nagai discs are combined to accurately match the gravitational accelerations produced by an exponential disc. We modelled the disc prior to ram pressure stripping to try to understand its response to ram pressure. Thus, we choose a disc mass of $\rm 4.5\times10^7\,M_\odot$ that consists of the currently detected $\hi$ disc mass ($\rm 3\times10^7\,M_\odot$) plus the $\hi$ gas detected in the tail ($\rm 1.5\times10^7\,M_\odot$). A radial exponential scale radius of 0.7 kpc is chosen, and we confirm that the resulting surface density profile is a reasonable match to the observed $\hi$ radial profile. 

To test the response to ram pressure, the restoring pressure from the self-gravity of the disc is compared at each location within the disc with the ram pressure strength from the cluster. If the ram pressure is stronger than the maximum restoring pressure found at a particular disc radius, then the gas at that radius can be assumed to be stripped. The restoring pressure that acts on the gas disc at each radius is calculated as the product of the gas disc surface density at a chosen cylindrical radius multiplied by the gravitational acceleration out of the plane of the disc (assuming face-on-ram pressure stripping). For the ram pressure from the cluster, we assume an orbital velocity of 500 $\rm km\,s^{-1}$ (the velocity difference of the galaxy with respect to the systemic velocity of the Virgo cluster’s brightest central galaxy, \citealt{2020A&A...635A.135Kashibadze}
), and a beta model for the Virgo ICM with $\rm \beta$=0.5, $\rm R_{core}$=50 kpc and $\rm \rho_0=2.0\times 10^{-26}$ $\rm g/cc$ (Model C1 from \citealt{2007MNRAS.380.1399Roediger})

\begin{figure}
\includegraphics[width=9cm]{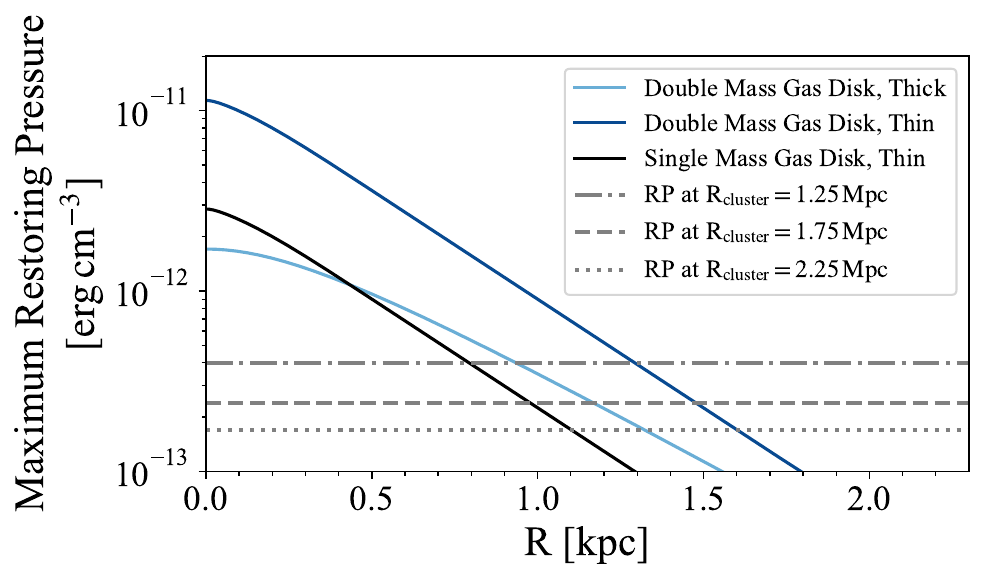}
\caption{Galactic radius vs the maximum restoring pressure by self-gravity of a gas disc. The solid black curve is for a thin gas disc with mass equal to that of AGC 226178, the solid dark blue curve for a thin gas disc mass twice that of AGC 226178, and the solid light blue curve for a thick gas disc of twice the mass of AGC 226178. The horizontal lines indicate ram pressure (RP) levels at three different cluster-centric distances from Virgo. The $\hi$ gas is stripped if the ram pressure is larger than the maximum restoring pressure. See Section \ref{sec:rps_test} for details.}
\label{fig: model}
\end{figure}

By assuming a thin disc, the restoring pressure from the self-gravity of the disc is maximized. In Fig. \ref{fig: model}, we plot the maximum restoring pressure as a function of the cylindrical disc radius. Since the projected distance of AGC 226178 from the cluster centre (1.25 Mpc) represents a lower limit to its true three-dimensional distance, we calculate the ram pressure assuming cluster-centric distances of 1.25, 1.75, and 2.25 Mpc, respectively. As can be seen, the restoring pressure from the $\rm 4.5\times10^7\,M_\odot$ disc is insufficient to maintain an extended $\hi$ distribution as observed in AGC 226178 ($>$ 2 kpc in radius) for cluster-centric distances up to 2.25 Mpc. To maintain its current $\hi$ size, the cloud has be to at a cluster-centric distance far beyond 3$\times$ the virial radius of the Virgo cluster ($\sim$ 1.0 Mpc). 
We also test a gas disc that is double the observed mass of AGC 226178 ($\rm 9.0\times10^7\,M_\odot$). This allows a larger and more extended gas disc to survive the ram pressure. But to allow for a $\hi$ gas disc as extended as AGC 226178, the cluster-centric distance should be at least three times the virial radius of Virgo.

Our toy model presented above neglects the potentially relevant thermal (or confinement) pressure exerted by the ICM. However, several studies suggest that confinement pressure from the ICM could help stabilize such systems. \citet{Burkhart2016} proposed that $\hi$ clouds may reach pressure equilibrium with the external ICM, and hydrodynamical simulations by \citet{Bellazzini2018} and \citet{2020MNRAS.499.5873Calura} further demonstrated that, under confinement pressure, $\hi$ clouds may survive for timescales of $\gtrsim$1 Gyr. While these simulations provide valuable insights into the stability of these intriguing systems, they remain far from fully realistic. For example, they assumed a low to moderate velocityaa of the clouds relative to the ICM, such that the ram pressure—a key environmental factor shown to generally dominate over confinement pressure in galaxy clusters \citep[][]{Bahe2012}—is less important than thermal pressure. For AGC 226178, if assuming a Virgo-centric distance of 1.7 Mpc (where the average density of the ICM is $\sim$ 10$^{-28}$ g cm$^{-3}$), a mean molecular weight of 0.6 for the hot ICM, an ICM temperature of 10$^{7}$ K, and a relative velocity of 500 km s$^{-1}$, we estimate a ram pressure of 2.5$\times$10$^{-13}$ erg cm$^{-3}$ and a thermal confinement pressure of 1.4$\times$10$^{-13}$ erg cm$^{-3}$, suggesting a ram-to-thermal pressure ratio of 1.8. The three-dimensional relative velocity is likely larger than 500 km s$^{-1}$, so the pressure ratio should be even higher. Therefore, while thermal confinement may contribute to the apparent stability of AGC 226178, it is unlikely to be the dominant factor. If the object resides within the cluster environment, its continued existence would therefore require a substantial reservoir of unaccounted baryonic mass.

Recent wide-field searches for isolated blue stellar blobs towards the Virgo cluster by \citet{2025ApJ...983....2Dey} suggest that such stellar systems are relatively common and tend to concentrate in regions with significant X-ray emission from the ICM, while generally avoiding the central core. This apparent distribution was interpreted as evidence that these objects have passed pericentre and are now moving outwards along the elongated orbits. However, a more natural explanation is that many of these systems may actually lie far beyond the cluster in three-dimensional space but are projected onto the Virgo region. Indeed, phase-space analyses show that the interloper fraction increases significantly with projected radius from cluster centres, and many of the objects identified by \cite{2025ApJ...983....2Dey} fall into high-probability interloper zones—especially beyond 1–2 $R_{200}$ \citep{2010A&A...520A..30Mamon}. Therefore, while the location of AGC226178 is consistent with that of the \cite{2025ApJ...983....2Dey} sample, the uncertainty introduced by projection effects implies that its cluster membership cannot be definitively established from phase-space arguments alone. Nevertheless, if AGC226178 is indeed a Virgo member, environmental processes such as ram pressure stripping could plausibly contribute to the disturbed $\hi$ morphology.

We conclude that the long-term survivability of AGC~226178 depends critically on its baryonic mass and, plausibly, on the complex hydrodynamical interactions with the ICM. If the system is primarily composed of the observed neutral hydrogen gas, with negligible contribution from other baryonic components such as stars or molecular gas, then its dynamical-to-baryonic mass ratio implies that it is not gravitationally bound. In this case, AGC 226178 is likely located well outside the Virgo cluster core, where the low ICM density would allow for the preservation of its extended $\hi$ distribution. The system appears to be in the process of disintegration, as implied by the irregular morphology, $\hi$ tail-like substructures and the presence of nearby disjoint blue stellar systems (Section \ref{sec: lscalehi}). However, if a substantial fraction of the baryonic mass resides in unobserved components—most plausibly CO-dark molecular gas as implied in simulated tidal dwarf galaxies \citep[][]{2007Sci...316.1166Bournaud}—then the system may possess sufficient mass to maintain marginal dynamical stability in the Virgo cluster. Additionally, confinement pressure from the ICM may further contribute to the cloud’s extended survivability. 

\section{Conclusions}
We have presented a comprehensive analysis of $\hi$ 21-cm line properties of an almost dark galaxy candidate -- AGC 226178. The $\hi$ data includes VLA multi-configuration mapping and FAST OTF mapping, obtained recently as part of the AVID project. Our VLA observations resolve the $\hi$ distribution in AGC 226178 down to subkiloparsec scales, and our FAST observation achieve a 3$\sigma$ column density sensitivity of 4.8$\times$10$^{17}$ cm$^{-2}$ for a 20 km s$^{-1}$ linewidth. 

The FAST observation of AGC 226178 reveals a short tail extending towards the lower-velocity end in the south-west direction that points towards (in projection) the nearby dwarf galaxy VCC 2034, which has been considered the most probable point of origin of AGC 226178 in the recent literature. However, VCC 2034 exhibits a line-of-sight $\hi$ stripping signature that is inconsistent with an event that would have resulted in the formation of AGC 226178. The observed properties of the $\hi$ gas in AGC 226178 and its projected neighbouring galaxies rule out any physical association between them.

AGC 226178 has a peak $\hi$ surface density of $\gtrsim$ 6 $M_{\odot}$ pc$^{-2}$. The main body of this $\hi$ cloud has a deconvolved FWHM of 2.7 kpc. The $\hi$ distribution is highly fragmented and clumpy, with clumps associated with ongoing star formation exhibiting systematically higher surface densities ($\gtrsim$ 4 $M_{\odot}$ pc$^{-2}$) than those without star formation. The $\hi$ velocity field exhibits a velocity gradient but is dominated by turbulence and/or unresolved random motions. The resolved $\hi$ clumps follow tight size-mass and size-velocity dispersion relations, as expected for a turbulent fragmentation. $\hi$ clumps associated with star formation, as well as AGC 226178 and other almost dark $\hi$ clouds as a whole, broadly follow the $\hi$ mass-star formation rate relation of ordinary galaxies. 

The star formation in AGC 226178 was initiated within the past $\sim$ 100 Myr or so. The large $\hi$ mass makes it unlikely that the object was a former dark matter mini-halo. Our finding that AGC 226178’s baryonic kinetic energy significantly exceeds its gravitational binding energy suggests that the system is undergoing disintegration, in line with the presence of $\hi$ tail-like features and a nearby disjoint blue stellar system. A comparison between the gravitational anchoring force and the ram pressure exerted by the ICM suggests that the system would have been largely disrupted if located within the cluster environment. To maintain even marginal dynamical stability under such conditions, a significant reservoir of unaccounted baryonic mass—most plausibly in the form of dark molecular gas—would be required. Furthermore, confinement pressure from the hot ICM may also contribute to the system's continued existence within the cluster.

\begin{acknowledgements}
We sincerely appreciate the anonymous referee's highly favorable and constructive report on our work. We thank Dr. Zhijie Qu for kindly pointing out an error in Figure 12 in the first arXiv version of the manuscript.
We acknowledge support from the National Key Research and Development Program of China (grant No. 2023YFA1608100), and from the NSFC (grant Nos. 12122303, 11973039, and 12173079). This work is also supported by the China Manned Space Project (grant Nos. CMS-CSST-2021-B02, CSST-2021-A06 and CMS-CSST-2021-A07). We also acknowledge support from the CAS Pioneer
Hundred Talents Program, the Strategic Priority Research Program of Chinese Academy of Sciences (grant No. XDB 41000000), and the Cyrus Chun Ying Tang Foundations.

This work made use of data from FAST (Five-hundred-meter Aperture Spherical radio Telescope)(https://cstr.cn/31116.02.FAST). FAST is a Chinese national mega-science facility, operated by National Astronomical Observatories, Chinese Academy of Sciences. We thank Jing Wang and Dong Yang (Kavli Institute for Astronomy and Astrophysics, Peking University, Beijing 100871, People's Republic of China) for their assistance in developing the FAST observing strategy.

SHOH acknowledges a support from the National Research Foundation of Korea (NRF) grant funded by the Korea government (Ministry of Science and ICT: MSIT) (RS2022-00197685, RS-2023-00243222).

RS acknowledges financial support from FONDECYT Regular 2023 project No. 1230441 and also gratefully acknowledges financial support from ANID - MILENIO NCN2024$\_$112

\end{acknowledgements}

\bibliographystyle{aa} 
\bibliography{agc_refer} 

\begin{appendix}
\section{PV diagram of VLA observation for AGC 226178}

\begin{figure}[htp]
   \centering
   \includegraphics[width=9.1cm]{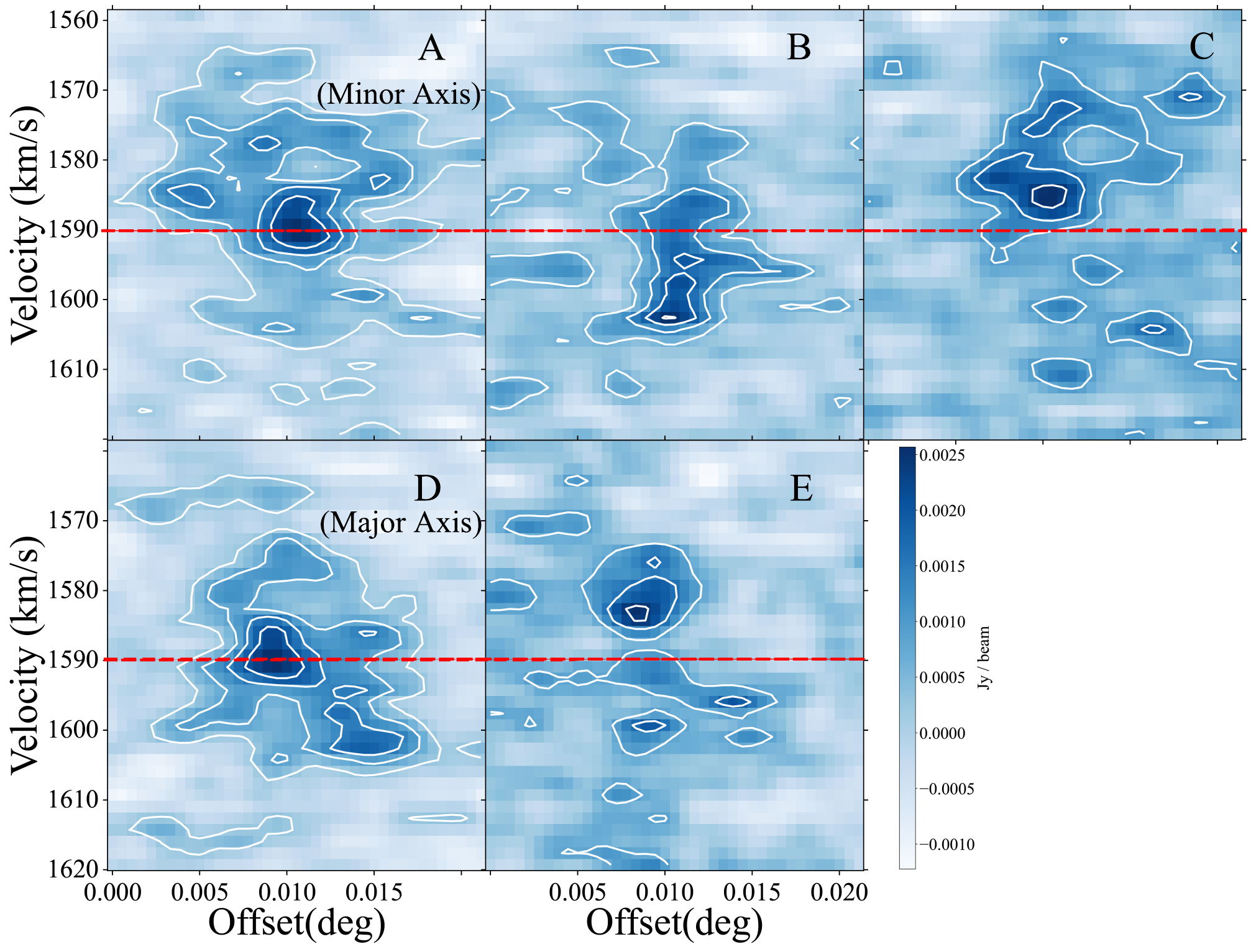}
   \caption{PV diagrams extracted along the directions indicated in the bottom middle panel of Fig.\ref{fig: VLA}, where slit A indicates the minor axis of AGC 226178 and slit D indicates the major axis. The contours represent 2, 4, 6, 8, and 12 times the RMS noise of 0.4 $\rm mJy\,beam^{-1}\,channel^{-1}$} 
   \label{fig: VLA_PV_model}
\end{figure}

To further explore the $\hi$ distribution across the velocity space, we extract position-velocity (PV) diagrams along five different directions indicated in the bottom middle panel of Fig. \ref{fig: VLA}, the PV diagrams are shown in Fig. \ref{fig: VLA_PV_model}. The extraction slits of the PV diagrams are chosen to be either along or parallel to the major and minor axes, where the major axis (slit D) matches the direction of the maximum $\hi$ velocity gradient whereas the minor axis (slit A) goes across the brightest $\hi$ peak. An apparent rotation-like velocity gradient is limited to the major axis. Moreover, the gradient appears to be driven by the two brightest clumps, with faint but notable $\hi$ gas in the `forbidden' quadrants for a rotation-dominated PV diagram.

\section{Properties of  $\hi$  clumps in AGC 226178.}

\begin{sidewaystable}
\caption{Properties of $\hi$ clumps in AGC 226178.}
\label{tab:clumps}
\centering
\begin{tabular}{lccccccc}
\hline\hline
Clump& RA & Dec  & $V_{\rm cen}$& Flux  & Spatial Size  & Velocity Width  & Associated $\hii$ Regions \\
ID &(deg) &(deg) & (km\,s$^{-1}$) &(mJy\,km\,s$^{-1}$)&(arcsec$\times$arcsec)&(km\,s$^{-1}$)& \\

\hline
1  & 191.67907 & 10.36797 & 1581.2 & 58.7 & 12.8$\times$18.1 & 7.2 & BC3s9, BC3s10, BC3s12, BC3s13, BC3s14 \\
2  & 191.67497 & 10.36574 & 1591.1 & 56.9 & 16.9$\times$18.1 & 9.0 & BC3s15, BC3s18, BC3s23 \\
3  & 191.67847 & 10.37008 & 1570.1 & 29.7 & 16.4$\times$12.9 & 6.2 & -- \\
4  & 191.67606 & 10.37303 & 1577.3 & 18.8 & 10.8$\times$11.3 & 5.0 & BC3s4, BC3s24W, BC3s26W \\
5  & 191.68145 & 10.36332 & 1573.7 & 12.7 & 7.4$\times$10.1  & 4.0 & BC3s19 \\
6  & 191.67465 & 10.36463 & 1578.1 & 13.5 & 8.1$\times$11.9  & 4.0 & BC3s20 \\
7  & 191.67906 & 10.36520 & 1568.9 & 18.1 & 14.3$\times$10.9 & 3.2 & -- \\
8  & 191.68031 & 10.36443 & 1591.1 & 4.4  & 5.7$\times$5.9   & 0.3 & -- \\
9  & 191.67911 & 10.37473 & 1574.3 & 8.1  & 9.5$\times$8.3   & 2.6 & -- \\
10 & 191.68112 & 10.36895 & 1589.5 & 8.3  & 6.4$\times$11.2  & 1.5 & -- \\
\hline
\end{tabular}
\tablefoot{The RA, Dec, and central velocity are centroid values returned by \texttt{CUPID}. Spatial sizes and velocity widths are Gaussian FWHM values, converted from standard deviations reported by \texttt{CUPID}. The quoted spatial sizes refer to the FWHM of the major and minor axes of the fitted 2D Gaussian. Associated $\hii$ region names are from \citet{2022ApJ...935...50Bellazzini}.}
\end{sidewaystable}
\end{appendix}

\end{document}